\input harvmac
\noblackbox
\input epsf
\def\abstract#1{
\vskip .5in\vfil\centerline
{\bf Abstract}\penalty1000
{{\smallskip\ifx\answ\bigans\leftskip 1pc \rightskip 1pc 
\else\leftskip 1pc \rightskip 1pc\fi
\noindent \abstractfont  \baselineskip=12pt
{#1} \smallskip}}
\penalty-1000}
%
\def\hth/#1#2#3#4#5#6#7{{\tt hep-th/#1#2#3#4#5#6#7}}
\def\nup#1({Nucl.\ Phys.\ $\us {B#1}$\ (}
\def\plt#1({Phys.\ Lett.\ $\us  {B#1}$\ (}
\def\cmp#1({Comm.\ Math.\ Phys.\ $\us  {#1}$\ (}
\def\prp#1({Phys.\ Rep.\ $\us  {#1}$\ (}
\def\prl#1({Phys.\ Rev.\ Lett.\ $\us  {#1}$\ (}
\def\prv#1({Phys.\ Rev.\ $\us  {#1}$\ (}
\def\mpl#1({Mod.\ Phys.\ Let.\ $\us  {A#1}$\ (}
\def\atmp#1({Adv.\ Theor.\ Math.\ Phys.\ $\us  {#1}$\ (}
\def\ijmp#1({Int.\ J.\ Mod.\ Phys.\ $\us{A#1}$\ (}
\def\jhep#1({JHEP\ $\us {#1}$\ (}

\def\subsubsec#1{\vskip 0.2cm \goodbreak \noindent {\it #1} \br}

\def\bb#1{{\bar{#1}}}

\def\cx#1{{\cal #1}}
\def\tx#1{{\tilde{#1}}}
\def\hx#1{{\hat{#1}}}

\def\us#1{\underline{#1}}
\def\fc#1#2{{#1\over #2}}
\def\frac#1#2{{#1\over #2}}

\def\br{\hfill\break}
\def\noi{\noindent}

\def\al{\alpha}\def\ga{\gamma}\def\om{\omega}
\def\p{\partial}\def\xit{\tilde{\xi}}
\lref\AFM{S.~Ferrara, R.~Minasian and A.~Sagnotti,
``Low-Energy Analysis of $M$ and $F$ Theories on Calabi-Yau Threefolds,''
Nucl.\ Phys.\ B {\bf 474}, 323 (1996)
[arXiv:hep-th/9604097].}
\lref\BMip{P. Berglund and P. Mayr, work in progress.}
\lref\AFT{R.~D'Auria, S.~Ferrara and M.~Trigiante,
``Homogeneous special manifolds, orientifolds and solvable coordinates,''
Nucl.\ Phys.\ B {\bf 693}, 261 (2004)
[arXiv:hep-th/0403204].}
\lref\pere{E.~Perevalov,
``On the hypermultiplet moduli space of heterotic compactifications with  small
instantons,''
arXiv:hep-th/9812253.}
\lref\Hit{N.~Hitchin,
``Generalized Calabi-Yau manifolds,''
Quart.\ J.\ Math.\ Oxford Ser.\  {\bf 54}, 281 (2003)
[arXiv:math.dg/0209099].}
\lref\Dol{I.~V.~Dolgachev,
``Mirror symmetry for lattice polarized K3 surfaces,''
arXiv:alg-geom/9502005.}
\lref\grassi{A.~Grassi,
``Divisors on elliptic Calabi-Yau four folds and the superpotential in F
theory. 1,''
J.\ Geom.\ Phys.\  {\bf 28}, 289 (1998) [arXiv:math.AG/9704008].}
\lref\Morr{D.~R.~Morrison,
``Mirror symmetry and the type II string,''
Nucl.\ Phys.\ Proc.\ Suppl.\  {\bf 46}, 146 (1996)
[arXiv:hep-th/9512016].}
\lref\GGC{D.~Cremades, L.~E.~Ibanez and F.~Marchesano,
``SUSY quivers, intersecting branes and the modest hierarchy problem,''
JHEP {\bf 0207}, 009 (2002)
[arXiv:hep-th/0201205]; D.~L\"ust, P.~Mayr, R.~Richter and S.~Stieberger,
``Scattering of gauge, matter, and moduli fields from intersecting branes,''
Nucl.\ Phys.\ B {\bf 696}, 205 (2004)
[arXiv:hep-th/0404134].}
\lref\TOR{E.~Witten,
``Phases of N = 2 theories in two dimensions,''
Nucl.\ Phys.\ B {\bf 403}, 159 (1993)
[arXiv:hep-th/9301042];\br
P.~S.~Aspinwall, B.~R.~Greene and D.~R.~Morrison,
``Calabi-Yau moduli space, mirror manifolds and spacetime topology  change in
string theory,''
Nucl.\ Phys.\ B {\bf 416}, 414 (1994)
[arXiv:hep-th/9309097].}
\lref\GLW{M.~Grana, J.~Louis and D.~Waldram, to appear.} 
\lref\HS{J.~A.~Harvey and A.~Strominger,
``The heterotic string is a soliton,''
Nucl.\ Phys.\ B {\bf 449}, 535 (1995)
[Erratum-ibid.\ B {\bf 458}, 456 (1996)]
[arXiv:hep-th/9504047].}
\lref\BaBe{V.~Balasubramanian and P.~Berglund,
``Stringy corrections to Kaehler potentials, SUSY breaking, and the
cosmological constant problem,''
JHEP {\bf 0411}, 085 (2004)
[arXiv:hep-th/0408054];\br
V.~Balasubramanian, P.~Berglund, J.~P.~Conlon and F.~Quevedo,
``Systematics of moduli stabilisation in Calabi-Yau flux compactifications,''
arXiv:hep-th/0502058.}
\lref\hpkatz{http://www.math.okstate.edu/~katz/CY/}
\lref\Mich{J.~Michelson,
``Compactifications of type IIB strings to four dimensions with  non-trivial
classical potential,''
Nucl.\ Phys.\ B {\bf 495}, 127 (1997)
[arXiv:hep-th/9610151];\br
G.~Dall'Agata,
``Type IIB 
supergravity compactified on a Calabi-Yau manifold with  H-fluxes,''
JHEP {\bf 0111}, 005 (2001)
[arXiv:hep-th/0107264].}
\lref\refcmap{S.~Cecotti, S.~Ferrara and L.~Girardello,
``Geometry Of Type Ii Superstrings And The Moduli Of Superconformal Field
Theories,''
Int.\ J.\ Mod.\ Phys.\ A {\bf 4}, 2475 (1989);\br
S.~Ferrara and S.~Sabharwal,
``Quaternionic Manifolds For Type Ii Superstring Vacua Of Calabi-Yau Spaces,''
Nucl.\ Phys.\ B {\bf 332}, 317 (1990).}
\lref\BM{P.~Berglund and P.~Mayr,
``Heterotic string/F-theory duality from mirror symmetry,''
Adv.\ Theor.\ Math.\ Phys.\  {\bf 2}, 1307 (1999)
[arXiv:hep-th/9811217].}
\lref\Witc{E.~Witten,
``Some comments on string dynamics,''
arXiv:hep-th/9507121.}
\lref\SWts{N.~Seiberg and E.~Witten,
``Comments on String Dynamics in Six Dimensions,''
Nucl.\ Phys.\ B {\bf 471}, 121 (1996)
[arXiv:hep-th/9603003].}
\lref\Douglasbrane{M.~R.~Douglas,
``Branes within branes,''
arXiv:hep-th/9512077.}
\lref\IBW{P.~G.~Camara, L.~E.~Ibanez and A.~M.~Uranga,
``Flux-induced SUSY-breaking soft terms,''
Nucl.\ Phys.\ B {\bf 689}, 195 (2004)
[arXiv:hep-th/0311241];\br
P.~G.~Camara, L.~E.~Ibanez and A.~M.~Uranga,
``Flux-induced SUSY-breaking soft terms on D7-D3 brane systems,''
Nucl.\ Phys.\ B {\bf 708}, 268 (2005)
[arXiv:hep-th/0408036];\br
A.~Font and L.~E.~Ibanez,
``SUSY-breaking soft terms in a MSSM magnetized D7-brane model,''
arXiv:hep-th/0412150; }
\lref\ADFT{C.~Angelantonj, R.~D'Auria, S.~Ferrara and M.~Trigiante,
``K3 x T**2/Z(2) orientifolds with fluxes, open string moduli and  critical
points,''
Phys.\ Lett.\ B {\bf 583}, 331 (2004)
[arXiv:hep-th/0312019].}
\lref\IBWii{D.~L\"ust, S.~Reffert and S.~Stieberger,
``Flux-induced soft supersymmetry breaking in chiral type IIb orientifolds with
D3/D7-branes,''
Nucl.\ Phys.\ B {\bf 706}, 3 (2005)
[arXiv:hep-th/0406092];\br
D.~L\"ust, S.~Reffert and S.~Stieberger,
``MSSM with soft SUSY breaking terms from D7-branes with fluxes,''
arXiv:hep-th/0410074;\br
D.~L\"ust, P.~Mayr, S.~Reffert and S.~Stieberger,
``F-theory flux, destabilization of orientifolds and soft terms on D7-branes,''
arXiv:hep-th/0501139; 
for a review and more references see also D.~L\"ust,
``Intersecting brane worlds: A path to the standard model?,''
Class.\ Quant.\ Grav.\  {\bf 21}, S1399 (2004)
[arXiv:hep-th/0401156].}

\lref\Ruben{S.~Fidanza, R.~Minasian and A.~Tomasiello,
``Mirror symmetric SU(3)-structure manifolds with NS fluxes,''
Commun.\ Math.\ Phys.\  {\bf 254}, 401 (2005)
[arXiv:hep-th/0311122];\br
M.~Grana, R.~Minasian, M.~Petrini and A.~Tomasiello,
``Supersymmetric backgrounds from generalized Calabi-Yau manifolds,''
JHEP {\bf 0408}, 046 (2004)
[arXiv:hep-th/0406137].}
\lref\MNVW{J.~A.~Minahan, D.~Nemeschansky, C.~Vafa and N.~P.~Warner,
``E-strings and N = 4 topological Yang-Mills theories,''
Nucl.\ Phys.\ B {\bf 527}, 581 (1998)
[arXiv:hep-th/9802168].}
\lref\VWx{C.~Vafa and E.~Witten,
``A Strong coupling test of S duality,''
Nucl.\ Phys.\ B {\bf 431}, 3 (1994)
[arXiv:hep-th/9408074].}
\lref\AHM{P.~S.~Aspinwall,
``Aspects of the hypermultiplet moduli space in string duality,''
JHEP {\bf 9804}, 019 (1998)
[arXiv:hep-th/9802194].}
\lref\SDii{R.~Friedman, J.~Morgan and E.~Witten,
``Vector bundles and F theory,''
Commun.\ Math.\ Phys.\  {\bf 187}, 679 (1997)
[arXiv:hep-th/9701162];\br
P.~S.~Aspinwall and D.~R.~Morrison,
``Point-like instantons on K3 orbifolds,''
Nucl.\ Phys.\ B {\bf 503}, 533 (1997)
[arXiv:hep-th/9705104].}
\lref\PMade{P.~Mayr,
``Conformal field theories on K3 and three-dimensional gauge theories,''
JHEP {\bf 0008}, 042 (2000)
[arXiv:hep-th/9910268].}
\lref\AD{S.~Ashok and M.~R.~Douglas,
``Counting flux vacua,''
JHEP {\bf 0401}, 060 (2004)
[arXiv:hep-th/0307049].}
\lref\CF{A.~Ceresole, R.~D'Auria, S.~Ferrara and A.~Van Proeyen,
``Duality transformations in supersymmetric Yang-Mills theories coupled to
supergravity,''
Nucl.\ Phys.\ B {\bf 444}, 92 (1995)
[arXiv:hep-th/9502072].}
\lref\GLMW{
S.~Gurrieri, J.~Louis, A.~Micu and D.~Waldram,
``Mirror symmetry in generalized Calabi-Yau compactifications,''
Nucl.\ Phys.\ B {\bf 654}, 61 (2003)
[arXiv:hep-th/0211102];\br
S.~Gurrieri and A.~Micu,
``Type IIB theory on half-flat manifolds,''
Class.\ Quant.\ Grav.\  {\bf 20}, 2181 (2003)
[arXiv:hep-th/0212278].}
\lref\LGM{
J.~Louis and A.~Micu,
``Type II theories compactified on Calabi-Yau threefolds in the presence  of
background fluxes,''
Nucl.\ Phys.\ B {\bf 635}, 395 (2002)
[arXiv:hep-th/0202168].}
\lref\KaP{
S.~Kachru and A.~K.~Kashani-Poor,
``Moduli potentials in type IIA compactifications with RR and NS flux,''
arXiv:hep-th/0411279.}
\lref\NOV{A.~Neitzke and C.~Vafa,
``N = 2 strings and the twistorial Calabi-Yau,''
arXiv:hep-th/0402128.\br
N.~Nekrasov, H.~Ooguri and C.~Vafa,
``S-duality and topological strings,''
JHEP {\bf 0410}, 009 (2004)
[arXiv:hep-th/0403167].}
\lref\FerH{R.~D'Auria, S.~Ferrara, M.~Trigiante and S.~Vaula,
``Scalar potential for the gauged Heisenberg algebra and a non-polynomial
antisymmetric tensor theory,''
Phys.\ Lett.\ B {\bf 610}, 270 (2005)
[arXiv:hep-th/0412063];\br
R.~D'Auria, S.~Ferrara, M.~Trigiante and S.~Vaula,
``Gauging the Heisenberg algebra of special quaternionic manifolds,''
Phys.\ Lett.\ B {\bf 610}, 147 (2005)
[arXiv:hep-th/0410290].}
\lref\Silv{A.~Saltman and E.~Silverstein,
``The scaling of the no-scale potential and de Sitter model building,''
JHEP {\bf 0411}, 066 (2004)
[arXiv:hep-th/0402135].}
\lref\Banks{T.~Banks,
``Landskepticism or why effective potentials don't count string models,''
arXiv:hep-th/0412129.}
\lref\BKKM{P.~Berglund, S.~Katz, A.~Klemm and P.~Mayr,
``New Higgs transitions between dual N = 2 string models,''
Nucl.\ Phys.\ B {\bf 483}, 209 (1997)
[arXiv:hep-th/9605154].}
\lref\AM{I.~Antoniadis and T.~Maillard,
``Moduli stabilization from magnetic fluxes in type I string theory,''
arXiv:hep-th/0412008.}
\lref\PMff{P.~Mayr,
``Mirror symmetry, N = 1 superpotentials and tensionless strings on  Calabi-Yau
four-folds,''
Nucl.\ Phys.\ B {\bf 494}, 489 (1997)
[arXiv:hep-th/9610162].}
\lref\HM{J.~A.~Harvey and G.~W.~Moore,
  ``Algebras, BPS States, and Strings,''
  Nucl.\ Phys.\ B {\bf 463}, 315 (1996)
  [arXiv:hep-th/9510182].}
\lref\Witpt{E.~Witten,
``Phase Transitions In M-Theory And F-Theory,''
Nucl.\ Phys.\ B {\bf 471}, 195 (1996)
[arXiv:hep-th/9603150].}
\lref\KMV{A.~Klemm, P.~Mayr and C.~Vafa,
``BPS states of exceptional non-critical strings,''
arXiv:hep-th/9607139.}
\lref\NL{A.~E.~Lawrence and N.~Nekrasov,
``Instanton sums and five-dimensional gauge theories,''
Nucl.\ Phys.\ B {\bf 513}, 239 (1998)
[arXiv:hep-th/9706025].}
\lref\DougI{M.~R.~Douglas and M.~Li,
``D-Brane Realization of N=2 Super Yang-Mills Theory in Four Dimensions,''
arXiv:hep-th/9604041.}
\lref\KLM{A.~Klemm, W.~Lerche and P.~Mayr,
``K3 Fibrations and heterotic type II string duality,''
Phys.\ Lett.\ B {\bf 357}, 313 (1995)
[arXiv:hep-th/9506112].}
\lref\BS{M.~Bershadsky and V.~Sadov,
``F-theory on K3 x K3 and instantons on 7-branes,''
Nucl.\ Phys.\ B {\bf 510}, 232 (1998)
[arXiv:hep-th/9703194].}
\lref\AL{P.~S.~Aspinwall and J.~Louis,
``On the Ubiquity of K3 Fibrations in String Duality,''
Phys.\ Lett.\ B {\bf 369}, 233 (1996)
[arXiv:hep-th/9510234].}
\lref\mirrlec{S.~Hosono, A.~Klemm and S.~Theisen,
``Lectures on mirror symmetry,''
arXiv:hep-th/9403096.}
\lref\Rtopdata{See e.g. P.~Candelas, X.~De La Ossa, A.~Font, S.~Katz and D.~R.~Morrison,
``Mirror symmetry for two parameter models. I,''
Nucl.\ Phys.\ B {\bf 416}, 481 (1994)
[arXiv:hep-th/9308083]; S.~Hosono, A.~Klemm, S.~Theisen and S.~T.~Yau,
``Mirror symmetry, mirror map and applications to Calabi-Yau hypersurfaces,''
Commun.\ Math.\ Phys.\  {\bf 167}, 301 (1995)
[arXiv:hep-th/9308122]; P.~Mayr,
``Phases of supersymmetric D-branes on Kaehler manifolds and the McKay
correspondence,''
JHEP {\bf 0101}, 018 (2001)
[arXiv:hep-th/0010223].}
\lref\refsin{A.~Klemm and P.~Mayr,
``Strong Coupling Singularities and Non-abelian Gauge Symmetries in $N=2$
String Theory,''
Nucl.\ Phys.\ B {\bf 469}, 37 (1996)
[arXiv:hep-th/9601014];\br
S.~Katz, D.~R.~Morrison and M.~Ronen Plesser,
``Enhanced Gauge Symmetry in Type II String Theory,''
Nucl.\ Phys.\ B {\bf 477}, 105 (1996)
[arXiv:hep-th/9601108].}
\lref\KPT{R.~Kallosh, A.~K.~Kashani-Poor and A.~Tomasiello,
``Counting fermionic zero modes on M5 with fluxes,''
arXiv:hep-th/0503138.}
\lref\STR{A.~Strominger,
``Massless black holes and conifolds in string theory,''
Nucl.\ Phys.\ B {\bf 451}, 96 (1995)
[arXiv:hep-th/9504090].}
\lref\Witfb{E.~Witten,
``Five-brane effective action in M-theory,''
J.\ Geom.\ Phys.\  {\bf 22}, 103 (1997)
[arXiv:hep-th/9610234].}
\lref\Ganor{O.~J.~Ganor,
``A note on zeroes of superpotentials in F-theory,''
Nucl.\ Phys.\ B {\bf 499}, 55 (1997)
[arXiv:hep-th/9612077].}
\lref\PPl{S.~Kachru, A.~Klemm, W.~Lerche, P.~Mayr and C.~Vafa,
``Nonperturbative results on the point particle limit of N=2 heterotic string
compactifications,''
Nucl.\ Phys.\ B {\bf 459}, 537 (1996)
[arXiv:hep-th/9508155].}
\lref\DK{F.~Denef, M.~R.~Douglas, B.~Florea, A.~Grassi and S.~Kachru,
``Fixing all moduli in a simple F-theory compactification,''
arXiv:hep-th/0503124.}
\lref\vafaln{C.~Vafa,
``Superstrings and topological strings at large N,''
J.\ Math.\ Phys.\  {\bf 42}, 2798 (2001)
[arXiv:hep-th/0008142].}
\lref\DMW{M.~J.~Duff, R.~Minasian and E.~Witten,
``Evidence for Heterotic/Heterotic Duality,''
Nucl.\ Phys.\ B {\bf 465}, 413 (1996)
[arXiv:hep-th/9601036].}
\lref\PS{J.~Polchinski and A.~Strominger,
``New Vacua for Type II String Theory,''
Phys.\ Lett.\ B {\bf 388}, 736 (1996)
[arXiv:hep-th/9510227].}
\lref\aspe{A.~Sen,
``A note on enhanced gauge symmetries in M- and string theory,''
JHEP {\bf 9709}, 001 (1997)
[arXiv:hep-th/9707123];\br
P.~S.~Aspinwall,
``K3 surfaces and string duality,''
arXiv:hep-th/9611137.}
\lref\CGP{S.~Cecotti, L.~Girardello and M.~Porrati,
``Constraints On Partial Superhiggs,''
Nucl.\ Phys.\ B {\bf 268}, 295 (1986).}
\lref\Grana{M.~Grana,
``D3-brane action in a supergravity background: The fermionic story,''
Phys.\ Rev.\ D {\bf 66}, 045014 (2002)
[arXiv:hep-th/0202118].}
\lref\TTii{P.~K.~Tripathy and S.~P.~Trivedi,
``D3 brane action and fermion zero modes in presence of background flux,''
arXiv:hep-th/0503072.}
\lref\Saulina{N.~Saulina,
``Topological constraints on stabilized flux vacua,''
arXiv:hep-th/0503125.}
\lref\CFNOP{K.~Choi, A.~Falkowski, H.~P.~Nilles, M.~Olechowski and S.~Pokorski,
``Stability of flux compactifications and the pattern of supersymmetry
breaking,''
JHEP {\bf 0411}, 076 (2004)
[arXiv:hep-th/0411066].}
\lref\TriTri{P.~K.~Tripathy and S.~P.~Trivedi,
``Compactification with flux on K3 and tori,''
JHEP {\bf 0303}, 028 (2003)
[arXiv:hep-th/0301139].}
\lref\SW{N.~Seiberg and E.~Witten,
``Electric - magnetic duality, monopole condensation, and confinement in N=2
supersymmetric Yang-Mills theory,''
Nucl.\ Phys.\ B {\bf 426}, 19 (1994)
[Erratum-ibid.\ B {\bf 430}, 485 (1994)]
[arXiv:hep-th/9407087].}
\lref\Sen{A.~Sen,
``F-theory and Orientifolds,''
Nucl.\ Phys.\ B {\bf 475}, 562 (1996)
[arXiv:hep-th/9605150].}
\lref\BDS{T.~Banks, M.~R.~Douglas and N.~Seiberg,
``Probing F-theory with branes,''
Phys.\ Lett.\ B {\bf 387}, 278 (1996)
[arXiv:hep-th/9605199].}
\lref\VW{C.~Vafa and E.~Witten,
``Dual string pairs with N = 1 and N = 2 supersymmetry in four  dimensions,''
Nucl.\ Phys.\ Proc.\ Suppl.\  {\bf 46}, 225 (1996)
[arXiv:hep-th/9507050].}
\lref\MV{D.~R.~Morrison and C.~Vafa,
``Compactifications of F-Theory on Calabi--Yau Threefolds -- I,''
Nucl.\ Phys.\ B {\bf 473}, 74 (1996);
``Compactifications of F-Theory on Calabi--Yau Threefolds -- II,''
Nucl.\ Phys.\ B {\bf 476}, 437 (1996)
[arXiv:hep-th/9603161].}
\lref\BKQ{C.~P.~Burgess, R.~Kallosh and F.~Quevedo,
``de Sitter string vacua from supersymmetric D-terms,''
JHEP {\bf 0310}, 056 (2003)
[arXiv:hep-th/0309187].}
\lref\RS{D.~Robbins and S.~Sethi,
``A barren landscape,''
Phys.\ Rev.\ D {\bf 71}, 046008 (2005)
[arXiv:hep-th/0405011].}
\lref\DDF{F.~Denef, M.~R.~Douglas and B.~Florea,
``Building a better racetrack,''
JHEP {\bf 0406}, 034 (2004)
[arXiv:hep-th/0404257].}
\lref\LST{M.~Grana, T.~W.~Grimm, H.~Jockers and J.~Louis,
``Soft supersymmetry breaking in Calabi-Yau orientifolds with D-branes  and
fluxes,''
Nucl.\ Phys.\ B {\bf 690}, 21 (2004)
[arXiv:hep-th/0312232];\br
H.~Jockers and J.~Louis,
``The effective action of D7-branes in N = 1 Calabi-Yau orientifolds,''
Nucl.\ Phys.\ B {\bf 705}, 167 (2005)
[arXiv:hep-th/0409098].}
\lref\LJ{H.~Jockers and J.~Louis,
``D-terms and F-terms from D7-brane fluxes,''
arXiv:hep-th/0502059.}
\lref\LSTY{J.~Louis, J.~Sonnenschein, S.~Theisen and S.~Yankielowicz,
``Non-perturbative properties of heterotic string vacua compactified on  K3 x
T**2,''
Nucl.\ Phys.\ B {\bf 480}, 185 (1996)
[arXiv:hep-th/9606049].}
\lref\SVW{S.~Sethi, C.~Vafa and E.~Witten,
``Constraints on low-dimensional string compactifications,''
Nucl.\ Phys.\ B {\bf 480}, 213 (1996)
[arXiv:hep-th/9606122].}
\lref\GKTT{L.~G\"orlich, S.~Kachru, P.~K.~Tripathy and S.~P.~Trivedi,
``Gaugino condensation and nonperturbative superpotentials in flux
compactifications,''
arXiv:hep-th/0407130.}
\lref\TV{T.~R.~Taylor and C.~Vafa,
``RR flux on Calabi-Yau and partial supersymmetry breaking,''
Phys.\ Lett.\ B {\bf 474}, 130 (2000)
[arXiv:hep-th/9912152].}
\lref\PMssb{P.~Mayr,
``On supersymmetry breaking in string theory and its realization in brane
worlds,''
Nucl.\ Phys.\ B {\bf 593}, 99 (2001)
[arXiv:hep-th/0003198].}
\lref\CIV{F.~Cachazo, K.~A.~Intriligator and C.~Vafa,
``A large N duality via a geometric transition,''
Nucl.\ Phys.\ B {\bf 603}, 3 (2001)
[arXiv:hep-th/0103067].}
\lref\KV{S.~Ferrara, J.~A.~Harvey, A.~Strominger and C.~Vafa,
``Second quantized mirror symmetry,''
Phys.\ Lett.\ B {\bf 361}, 59 (1995)
[arXiv:hep-th/9505162];\br
S.~Kachru and C.~Vafa,
``Exact results for N=2 compactifications of heterotic strings,''
Nucl.\ Phys.\ B {\bf 450}, 69 (1995)
[arXiv:hep-th/9505105].}
\lref\KKLT{S.~Kachru, R.~Kallosh, A.~Linde and S.~P.~Trivedi,
``De Sitter vacua in string theory,''
Phys.\ Rev.\ D {\bf 68}, 046005 (2003)
[arXiv:hep-th/0301240].}
\lref\Witsp{E.~Witten,
``Non-Perturbative Superpotentials In String Theory,''
Nucl.\ Phys.\ B {\bf 474}, 343 (1996)
[arXiv:hep-th/9604030].}
\lref\ADFL{L.~Andrianopoli, R.~D'Auria, S.~Ferrara and M.~A.~Lledo,
``4-D gauged supergravity analysis of type IIB vacua on $K3 \times T^2/\IZ_2$,''
JHEP {\bf 0303}, 044 (2003)
[arXiv:hep-th/0302174].}
\lref\vafaf{C.~Vafa,
``Evidence for F-Theory,''
Nucl.\ Phys.\ B {\bf 469}, 403 (1996)
[arXiv:hep-th/9602022].}
\lref\FF{D.~L\"ust, P.~Mayr, S.~Reffert and S.~Stieberger,
``F-theory flux, destabilization of orientifolds and soft terms on D7-branes,''
arXiv:hep-th/0501139.}
\lref\GVW{
S.~Gukov, C.~Vafa and E.~Witten,
``CFT's from Calabi-Yau four-folds,''
Nucl.\ Phys.\ B {\bf 584}, 69 (2000)
[Erratum-ibid.\ B {\bf 608}, 477 (2001)]
[arXiv:hep-th/9906070].}
\lref\DRS{
K.~Dasgupta, G.~Rajesh and S.~Sethi,
``M theory, orientifolds and G-flux,''
JHEP {\bf 9908}, 023 (1999)
[arXiv:hep-th/9908088].}
\lref\GKP{S.~B.~Giddings, S.~Kachru and J.~Polchinski,
``Hierarchies from fluxes in string compactifications,''
Phys.\ Rev.\ D {\bf 66}, 106006 (2002)
[arXiv:hep-th/0105097].}
\lref\thebible{L.~Andrianopoli et al.,
``N = 2 supergravity and N = 2 super Yang-Mills theory on general scalar
manifolds: Symplectic covariance, gaugings and the momentum map,''
J.\ Geom.\ Phys.\  {\bf 23}, 111 (1997)
[arXiv:hep-th/9605032].}
\def\CY{Calabi--Yau}
\def\ofi{$T^2/\ZZ_2 \times$ K3}

\def\g{\underline{\rm{G}}}

\def\kf{\tx X_V}\def\kb{X_H}

\def\h{\fc{1}{2}}
\def\Pit{\tilde{\Pi}}
\def\La{\Lambda}\def\Om{\Omega}\def\Si{\Sigma}
\def\la{\lambda}
\def\hide#1{}
\def\IZ{{\bf Z}}\def\IP{{\bf P}}\def\IC{{\bf C}}
\def\Im#1{{\rm Im}\, #1}

\def\ktk{K3~$\times$~K3}
\def\etat{\tilde{\eta}}

\def\ofi{$T^2/\IZ_2 \times$ K3}
\def\TT{\rho}\def\SS{S_{II}}\def\Wi{W_{inst}}
\def\TTb{\bar{\rho}}
\def\qtt{Q}\def\MM{\cx M^\flat}
\def\BB{B}\def\Zh{\hat{Z}}\def\Zb{\breve{Z}}\def\gah{\hx \gamma}\def\Zsd{Z_{s.d.}}
\def\qh{\hx q}\def\tb{\bar{t}}
\def\S{$S$}\def\ffi{w_{inst}}\def\Fc{\cx F_{cubic}}\def\Sp{{\rm Sp}}
\vskip-2cm
\Title{\vbox{
\rightline{\vbox{\baselineskip12pt\hbox{LMU-ASC 28/05, UNH-05-02}
                            \hbox{hep-th/0504058}}}\vskip0.5cm}}
{Non-Perturbative Superpotentials in F-theory}
\vskip -0.8cm
\centerline{\titlefont and String Duality} 
\abstractfont

\vskip 1.3cm
\centerline{P. Berglund$^1$ and P. Mayr$^2$ }
\vskip 0.3cm
\centerline{$^1$ \it Department of Physics, University of New Hampshire,}
\centerline{\it Durham, NH 03824, USA}
\centerline{$^2$ \it Arnold--Sommerfeld--Center for Theoretical Physics,}
\centerline{\it Theresienstrasse 37, D-80333 Munich, Germany}
\vskip 0.5cm
\abstract{%
We use open-closed string duality between F-theory on \ktk\ and type II strings on 
CY manifolds without branes to study non-perturbative superpotentials in generalized flux
compactifications. On the F-theory side we obtain the full flux potential including
D3-instanton contributions and show that it leads to an explicit and simple realization
of the three ingredients of the KKLT model for stringy dS vacua. The D3-instanton
contribution is highly non-trivial, can be systematically computed including the determinant factors
and demonstrates that a particular flux lifts very effectively zero modes on the instanton.
On the closed string side, we describe a generalization of the Gukov-Vafa-Witten
superpotential for type II strings on generalized CY manifolds, depending on all moduli multiplets.
}
\Date{\vbox{\hskip-15pt April 2005 \hskip70pt 
$^1${\ninerm per.berglund@unh.edu}\hskip10pt 
$^2${\ninerm mayr@theorie.physik.uni-muenchen.de}}}
\goodbreak

\leftskip -0pc \rightskip 0pc 
\baselineskip 12pt

\newsec{Introduction}
One of the central questions in string theory is 
regarding 
the existence
and viability of ``semi-realistic'' 4d ground states. Much effort has been put
recently in the study of the vacuum structure of effective potentials in flux
compactifications. In particular ref.\KKLT\ (KKLT) outlined a qualitative picture of
how 4d meta-stable de Sitter vacua may arise in flux compactifications in string theory.

The method based on effective potentials is not undisputed \Banks\ and 
indeed this line of research is more experimental in spirit rather
than deriving properties of the string vacuum structure from first principles. On a more modest level, 
one would like to see whether the assumptions in the KKLT ansatz can be
met naturally and explicitly. There are essentially three basic ingredients: the classical
flux potential fixing complex structure moduli, a non-perturbative contribution
fixing K\"ahler moduli and a positive energy contribution that lifts the vacuum to de Sitter. 

In the KKLT ansatz, the non-perturbative part arises from (possibly ``fractional'')
D3-instantons, and these corrections to the potential have received much interest recently. 
D3-instanton contributions to the
superpotential have been first studied in \Witsp, where a quite restrictive condition
on the compactification geometry was found in order that such corrections
arise. As a consequence, the superpotential tends to be simple, often too simple 
to allow for an interesting potential that fixes all (K\"ahler) moduli \DDF\RS. This seems 
to require a considerable effort and expertise to find working examples \DDF\DK\ and 
makes the proposal of KKLT somewhat unhandy. Moreover
the computation of the full moduli dependence of the D3-instantons involves 
a 1-loop determinant, which depends in an intricate way on the moduli and has not been 
computed in general, in addition to summing over all possible instantons by hand.

In this paper we describe a simple flux compactification of F-theory on \ktk, where all
three ingredients of the KKLT ansatz fall natural into place and the full moduli
dependence of the non-perturbative D3-instanton sum can be computed explicitly. 
One of the important messages is that the flux is very effective in 
relaxing the geometric condition found in \Witsp, and leads to a highly non-trivial
instanton sum in a geometry that would be too simple to generate a potential
without flux.\foot{The condition in \Witsp\ was formulated in the dual M-theory compactification and 
says that the arithmetic genus of a divisor that contributes an instanton to the superpotential 
is $\chi=1$. It was argued in \GKTT\ that the flux should change this condition.
We find that in the presence of flux there are contributions from divisors with $\chi=0$ and
$\chi=2$.}. In particular the classical flux potential and the instanton part 
are intimately related and can not be considered independently. The explicit results are obtained by a
string duality that relates the F-theory compactification to a closed string compactification of
type II strings on CY, and they provide a nice testing ground for the ideas of \KKLT.
The generation of new instanton corrections with a rich structure in the presence of flux suggests
that the situation described in the KKLT ansatz is 
more generic and widespread than perhaps suggested by previous studies as refs.\DDF\RS\DK.

Apart from giving an explicit  realization of the KKLT model, 
the simple F-theory example on \ktk\ turns out to have a surprisingly rich structure 
also in other respects. One is the effective gauged $\cx N=2$ supergravity 
for this compactification and the dual type II string on a CY. 
The F-theory fluxes come close to realize the
most general gaugings considered in the supergravity literature so far \FerH,
gauging a Heisenberg algebra of the dual quaternionic space for the hyper multiplets. 
This turns out to be related to the presence of a large lattice of 4d BPS domain walls. 
The superpotential of the theory can be written in terms of the central charges 
of these domain walls and leads to a natural
generalization of the Gukov-Vafa-Witten
superpotential \GVW\ for type II strings, which is related to the recently discovered
S-duality of topological strings \NOV.

The organization of this paper is as follows. In sect.~2 we discuss general aspects 
of D3-instantons with a focus on the simple example\foot{This is related to a type IIB 
orientifold in a certain limit in the well-known way \Sen. We will
also comment on more general geometries.} of F-theory compactified on
\ktk\ and give some general arguments for the expected form of the instanton corrections.
The non-trivial world-volume theory is important to determine the determinant and
multi-wrapping effects.

In sect.~3 we propose a non-perturbative formulation of the F-theory flux potential in an effective
supergravity, which predicts already to some extent the structure of the D3-instanton corrections, 
including the moduli dependence of the determinant. This potential reproduces all three
ingredients of the KKLT ansatz without further adaptations.

In sect.~4 we describe the basic duality of F-theory on \ktk\ and type II strings
on CY manifolds with a focus on the mapping of 
instantons in the different formulations. In sect.~5 we use this duality to 
study the D3-instanton contributions to the superpotential. Moreover we compute 
perturbative couplings of D3/D7-branes in the F-theory and discuss some aspects of the 
vacuum structure, such as the fixing of the moduli of space-filling D3-branes.

In sect.~6 we propose a generalization of the Gukov-Vafa-Witten superpotential for the 
dual type II compactifications on generalized manifolds, as a bilinear 
\eqn\wgeni{W\sim\sum (\int_{\Zh}\Omega \wedge \hx \ga^I) \ \g_{I\La} (\int_{Z} \Omega \wedge \ga^\La),}
in the 'quantum periods' of a pair $(Z,\Zh)$ of Calabi--Yau mirror manifolds. The duality 
implies that particular scalars of all  hypers can be charged under all $U(1)$ symmetries
gauged by the vectors, the existence of supersymmetric branches,  
a large set of new domain walls that are exchanged with D5 and NS5 branes under generalized T-dualities,
and a modified statistics of flux vacua.

In sect.~7 we conclude with a preliminary study of the vacuum structure. We find an effective enhancement mechanism for
the instanton corrections that relaxes a restrictive bound found in \KKLT\ and argue that
most of the 4d $\cx N=1$ 
supersymmetric ground states in the no-scale approximation are extinguished by the
instanton effects.

In the Appendix we collect some necessary computations on the geometry of the dual CY manifolds.
\vskip12pt

\noi
{\bf Note added:} While completing this work, three papers appeared that consider the effect of the 
flux from the point of the world-volume theory on the brane instanton \TTii\Saulina\KPT. 
Some of the instantons computed in this paper fall into the class discussed in these papers,
whereas others are of a more general type.
The results obtained in this paper are complementary, 
as we obtain the full instanton sums including the dependence
on the remaining moduli, multi-wrappings, non-isolated D3-instantons and other effects. On the other
hand a world-volume interpretation of these results is not at all trivial;
it would be interesting to recover some of the phenomena we find also 
from the world-volume perspective.

\newsec{D3-instantons: Structure and some physical arguments}
In this section we discuss some general aspects of M5/D3 instantons, with a focus on the concrete 
example \ktk. Taking into account the non-trivial world-volume structure of the 5-brane 
leads to a heuristic derivation of some of the later results.

The basic instanton in the 3d theory obtained
by compactifying M-theory on a 4-fold $X_4$ is a M5-brane wrapped on a divisor $D$ in $X_4$ \Witsp. If
$X_4$ has an elliptic fibration over a base $B_3$ 
one can take the small fiber limit of $X_4$ to obtain a 4d compactification
of F-theory on $X_4$ \vafaf\MV. It was further argued in \Witsp\ that the 4d instantons arise from vertical
divisors in this limit, that is divisors $D$ that map to a divisor $S$ in the base $B_3$. In 
F-theory, the dual of the M5-brane wrapped on $D$ is a D3-brane wrapped on $S$.

The contribution of a single D3-instanton to the superpotential is of the form
\eqn\Dinst{
W\sim f(...)\, e^{2\pi i \TT},\qquad\ \  \TT=\TT_1+i\TT_2=\int C_4 + i Vol(S),
}
where $C_4$ is the RR 4-form of type IIB. The 'coefficient function' $f(...)$ is 
a holomorphic section of a line bundle\foot{The full $\cx N=1$ superpotential $W$ is the section of 
a different line bundle, whose definition includes the moduli dependence from the instanton weight. } 
that depends on the other moduli in the problem, 
including the position of space-filling D3-branes. This function may in principle be computed
from a 1-loop computation on the world-volume of the underlying M5 brane \Witsp\Witfb\Ganor.

A contribution to the superpotential requires that the instanton supports two fermionic
zero modes. In the absence of flux, and for a smooth divisor $D$, the presence of
precisely two zero modes translates to the following necessary condition on the divisor $D$ \Witsp:
$$
\chi(D)=\sum_p h^{0,p}(D)=1.
$$ 
The above index counts the net number of fermionic zero modes weighted by their charge
under the circle  action defined in the normal bundle of $D$ in $X_4$. Fermions of opposite
$U(1)$ charge may get massive in pairs and thus the index is a lower bound on the number of 
zero modes. 

For $X_4=$ \ktk, the base of the fibration is $B_3=B_V\times \kb$, with $B_V$ the $\IP^1$ base of the
elliptic fibration $\kf\to B_V$.\foot{We denote the ``upper'', elliptically fibered K3  
by $\kf$ and the ``lower'' K3 by $\kb$. The subscript reminds of the fact that roughly
speaking the moduli of $\tx X_{V(ector)}$ and $X_{H(yper)}$ 
end up in 4d vector and hyper multiplets, respectively.
$X_{H(eterotic)}$ is also the K3 common with the heterotic dual introduced later on.}
There are three basic classes of divisors in $X_4$ that 
will be relevant in the following:\vskip10pt

\item{$A)$} A non-isolated, vertical divisor of $\chi=0$ that projects to the divisor $S=\kb$ in $B_3$.
\item{$B)$} A non-vertical divisor of $\chi=2$ of the form $C_1\times \kb$, with $C_1\in H_2(\kf)$.
\item{$C)$} A vertical divisor of $\chi=2$ that projects to a divisor $B_V\times C_1'$, with 
$C_1'\in H_2(\kb)$.

\vskip10pt 
\noi
The lower bound on the number of fermionic zero modes, provided by the index, is 0,4,4 in the 
three cases, respectively.
These divisors have naively the wrong number of zero modes to support D3-instanton  contributions to the 
superpotential, however cases $B)$ and $C)$ have the right number of zero modes 
to contribute to the K\"ahler potential.\foot{Our
statements on the non-vertical divisors of type $B)$ will be somewhat preliminary.}

The divisors of type $A)$ have $h^{0,p}=1$, $p=0,...,3$, corresponding to 8 possible fermionic zero modes 
and the situation is more complicated.
These divisors have moduli from the non-trivial third cohomology and the instanton contribution is an integral
over the moduli. Moreover the non-trivial
$H^3$ requires to sum over the gauge field configurations on the 5-brane 
world-volume, which leads to a moduli dependence in the form 
of a theta function \Witfb. We will be able to compute these effects explicitly and
see that these non-isolated divisors contribute to amplitudes as if they had four zero modes.

Moreover we will argue that upon adding flux in F-theory,
divisors of type $A)$ (and possibly $B)$) may contribute to the 4d superpotential.

Switching on flux in the compactification manifold has essentially two effects on the instantons. 
First, the flux may lift some of the zero modes such that the same instanton may
contribute to a different amplitude that can absorb less zero modes, such as a superpotential.
An important point is that since the instantons exist already in the parent theory
without flux, which has more supersymmetry in general, the contribution to the
superpotential can be often obtained by computing a different amplitude in the simpler
parent theory without flux. This has been heavily used in \TV\PMssb\vafaln\CIV.

The second effect of the flux is that it may generate a tadpole on the brane
and obstruct an instanton \DMW. This leads to further consistency conditions in flux compactifications,
which may interfere with the goal to fix all moduli.
A background flux may therefore generate new instanton corrections as well as 
obstruct instanton corrections, depending on the orientation
of the flux in $X_4$ relative to the brane wrapping on $D$.

We will determine the flux that is relevant for the appropriate lift of the
zero modes below; in particular for each instanton there is only one 
flux component that lifts the zero modes, namely the flux in the $U(1)$ under which
the instanton is magnetically charged. Let us for now assume that the flux is 
of the appropriate kind to reduce the
number of zero modes on the M5-brane wrapping $D$ to allow a contribution to the
superpotential. What will be the 
contribution of the instanton on $D$ to the 4d superpotential? The 3d BPS instanton is related to a 
4d magnetic monopole $M$, with a mass proportional to the action of the instanton. 
In F-theory, the 4d magnetic
monopole is given by a D5-brane stretching between two 7-branes and wrapped on $S$  \DougI.
As usual in string theory, the magnetic monopole represents a fundamental state in 
the appropriate regime of the moduli space, where the volume of the divisor,
$Vol(S)=\TT_2$, gets small, and the instanton weight \eqn\instw{Q=\exp(2\pi i \TT),} is 
$Q\sim 1$. It is not known at this point in which super multiplet 
the light monopole is, but in any case, it will lead to a logarithmic singularity
in the complexified gauge coupling, of the form
\eqn\effgc{
\tau_{eff}=\fc{\theta}{2\pi}+\fc{i}{g^2}=\tau_0+c\ \ln(1-Q)+...,
}
where $c$ is a constant proportional to the beta function coefficient of the monopole
multiplet. 

As explained above, the role of the flux is to lift zero modes and the same instanton
contributes already to a different amplitude in the parent theory without flux. 
In the present context, this is the amplitude in the parent theory which can absorb four zero modes, 
namely the prepotential $\cx F$. The prepotential determines also the gauge coupling \effgc\ and 
near the singularity one has approximately $\tau_{eff}=\p_\TT^2 \cx F$. The superpotential
is therefore closely related to $\cx F$ and to the gauge coupling. We will show that 
the relation between the gauge coupling and the  flux superpotential is simply%
\foot{Crudely speaking, the reduction of the power of the derivative reflects the reduced
number of zero modes, see \vafaln\ for a related discussion.}
\eqn\wrel{
W_{inst}(\TT)\sim\p_\TT \cx F = -{b(M)\over 2\pi i} \int \ln(1-Q)\,  d\TT = {b(M)\over (2\pi i)^2}\, \sum \fc{Q^n}{n^2},}
where $b(M)$ is the beta function coefficient of the monopole multiplet.
In the absence of further effects the above  potential has the typical decompactification
behavior, with $Q=0$ the only solution to $\p_\TT W=0$.

Eq.\wrel\ predicts that an infinite number of D3-branes from multi-wrappings on $S$  
contribute to the superpotential.
This does not follow immediately from the logarithmic divergence in eq.\effgc\ because of the 
extra terms hidden in the dots. 
However if the compactification is related to a M-theory compactification in 
5d, as is the case for the present compactification, the $Q$ expansion of the gauge coupling
is as in \effgc\ \NL. The reason is that the 4d instantons arise from word-lines of 5d particles and multiply wrapped
branes correspond to multiple instantons. These instantons then contribute to the superpotential as in \wrel.

Further details of the potential depend on other (massive) 
physical states in the theory, which also enter the dots in \effgc. These effects 
depend on the details of the full theory and can not be recovered from this
simple argument.\foot{In particular, D3-instantons of type $A)$ have moduli from the position
relative to the 7-branes and the full action gets contributions from various patches in the 
transverse D3-brane position.}  
One can also not deduce the ``constant'' $f$ in \Dinst, which depends
on the other moduli. 
These more detailed questions will be addressed in the full string compactification, which has much
more structure than the simple field theory picture sketched above. 

One important ingredient is 
that the M-theory 5-brane that describes the basic instanton of \Witsp, has itself a non-trivial
world-volume theory that may lead to a proliferation of instanton effects. In the present case
the instanton corrections get contributions from the massive states of a non-critical string,
of which the 4d monopole is the lightest state. The close connection between 
the 5-brane instanton corrections and non-critical strings was already noted in \PMff,
and holds also for the examples discussed in \DDF\DK. This will be further discussed in sect.~5.

One comment concerning the divisors in $B)$. The general argument of \Witsp, that
only vertical divisors can contribute in the 4d limit relies on the relation 
$$
\fc{1}{g_{3d}^2}=\fc{1}{g_{4d}^2}\ R
$$
where $R$ is the radius in the compactification from 4d to 3d. The volume of 
a brane wrapped on the elliptic fiber $E$ of volume $\sim 1/R$ compensates the radius dependent term, 
whereas branes not wrapped on $E$ die out in the 4d limit with a positive power of 
$e^{-R}$. However one can conceive a more complicated limit, where one also changes the 
complex structure of the elliptically fibered K3 such that a divisor of the type $B)$ has 
a size comparable to a vertical divisor.
This limit corresponds to  moduli near a non-Abelian gauge symmetry enhancement in 4d
and divisors of these type could play a role in reproducing the gauge theory instantons.

\newsec{Flux potential and the KKLT ansatz}
In this section we describe the four-dimensional flux potential for 
the F-theory compactification in the form that will be used for the computation by string duality.
Specifically we write the instanton corrected potential in terms of an
effective supergravity that may be used to relate the two dual theories.
The effective potential obtained in this way is precisely of the form of the KKLT
ansatz, without any further adaptations.

We will mainly consider the case of F-theory compactified on \ktk, although some of the
arguments generalize to other geometries.

\subsec{Classical potential}
The standard superpotential for the type IIB string compactified to 4d with bulk 3-form fluxes is \GVW\TV\
\eqn\spst{
W_F = \int \Omega \wedge G_3,
}
where $\Omega$ is the holomorphic $(3,0)$ form. Moreover $G_3=F_3-\SS H_3$, with $\SS$ the complex type IIB dilaton
and $F_3$ ($H_3$) the RR (NS) 3-form flux. In the case of a purely closed string compactification on 
a Calabi--Yau 3-fold $X$ the expression \spst\ is an {\it exact} formula, including an infinite series of instanton
corrections \TV\PMssb. One way to derive this fact is to consider the superpotential in terms
of the quantities of the underlying $\cx N=2$ effective supergravity for the compactification without flux. 
Under certain conditions it can be written as \PMssb
\eqn\spsugra{
W_{eff}=\sum_\La P^{hol}_\La\, X^\La-P^{\La}_{hol}F_\La,}
where $(X^\La,F_\La)$ is the upper half of the symplectic section for the
$\cx N=2$ vector multiplets in the effective
supergravity\foot{We refer to \thebible\ for background material and references
on the effective supergravity.}. 
Moreover the objects $(P^{hol}_\La,\, P_{hol}^\La)$ are holomorphic versions of the 
standard Killing prepotentials that describe the gauging in the supergravity induced
by the background flux.\foot{Explicitly, $P_{hol}^\La=e^{-K_H/2}P_\La$, with $K_H$ the 
K\"ahler potential and $P_\La$ the 
Killing prepotential for the hyper multiplets.} The second term corresponds to a gauging of the magnetic
fields and will be often set to zero in the following.

The objects in \spsugra\ are understood as the exact,
non-perturbative expressions including all quantum effects. The exact equality 
$W_{eff}=W_F$ follows from the identities
\eqn\clrel{
X^\La\sim\int_{\gamma_\La} \Om,\qquad P^{hol}_\La \sim \int_{\ga^*_\La} G_3.
}
where $\ga_\La\in H_3(X,\IZ)$ is a basis of 3-cycles, and $\ga_\La^*$ the cycle dual to $\ga_\La$.
The exactness of the formula \spst\ is due to the well-known fact that the geometric periods 
$\int \Om$ compute the full quantum corrected supergravity section $X^\La$ \KV. Indeed
the formula \spst\ has been derived in \TV\ as the exact BPS tension of a 5-brane domain wall 
wrapping a 3-cycle in the Calabi--Yau manifold $X$.

\subsubsec{Compactifications with background branes}
The same formula \spst\ for the superpotential has been obtained in \GKP\ for flux compactifications 
in type IIB orientifolds with background branes, from a compactification of 
the 10d supergravity. However this expression 
is on a very different footing than in the closed string case 
and not equal to the exact tension of a BPS brane. We will now sketch
how this potential relates to an exact 4d supergravity quantity as in eq.\spsugra, in the case of 
F-theory on \ktk, which reduces to the type IIB orientifold \ofi\ in a certain limit.
Later we will compute
the exact potential in a dual theory described by the same effective supergravity.

Note that the dilaton $\SS$, which enters through the flux $G_3$,
is in a vector multiplet in the orientifold compactification on \ofi.\foot{See \ADFL\ADFT\ for 
a discussion of the effective supergravity of this orientifold.}
Thus morally speaking the assignment of vector and hyper multiplets
to the flux $G_3$ and the period integrals $\int \Omega$ is
reversed w.r.t. to the closed string identification \clrel.
Moreover, the expression \spst\ can not be the full answer for the orientifold, since it does not depend on the 
open string degrees of freedom and the back reaction of the dilaton to the 7-branes. 
The contributions of the open string sector have been studied in \FF\ in F-theory, building on 
the work \GVW. The result is 
\eqn\wf{
W_F=\sum_{I,\La}\ \Pi^I(q^i)\, \g_{I\La}\, \Pit^\La(t_a),\qquad  I=1,...,22,\ \La=1,...,20.
}
The objects appearing in the above formula are the period integrals  $\Pi^I$ and $\Pit^\La$ of the holomorphic 
$(2,0)$-form on the two K3 factors and an integral matrix $\g_{I\La}$ that specifies the flux background. 
More precisely, $\Pit^\La(t_a)$ are the periods on the ``upper'' K3, denoted by $\kf$,
which is elliptically fibered with a zero size fiber.
These periods depend only on scalars  in 4d vector multiplets, denoted by $t_a$.\foot{As mentioned already,
the type IIB dilaton $\SS$ is one of these scalars.}
On the elliptic fibration $\kf$ there are 20 periods labeled by $\La$, and these correspond 
to the 20 $U(1)$ massless gauge fields from the 7-branes at a generic point in the moduli. 
On the other hand, $\Pi^I(q^i)$ are the periods on the ``lower'' K3, denoted by $\kb$, 
and they depend only on scalars $q^i$ in hyper multiplets. 

Finally, the vacuum expectation values of the 20 independent $U(1)$ field strengths $\cx F_\La$ on the
lower K3 $\kb$ are specified as 
\eqn\Fflux{
\cx F_\La=\eta^I\g_{I\La},
}
with $\{\eta^I\}$ a basis of $H^2(\kb,\IZ)$. This flux induces at the same time an 
important $D$-term potential that will be discussed in sect.~3.3.

The K3 periods $\Pi^I(q^i)$ and $\Pit^\La(t_a)$ depend quadratically on the 
physical scalar fields $q^i$ and $t_a$, respectively. The potential \wf\ is therefore a quartic polynomial 
of degree (2,2) in the moduli $(q^i,t_a)$. 
To be more explicit and to make contact with the orientifold, 
let us spell out the dependence of the superpotential
on the scalars $t_a$. A certain parametrization of the period vector is
\eqn\pvkv{
\Pit^\La(t_a)=(1,-U  \SS+\fc{1}{2} C_aC_a,\SS,U,C_a)^T,\qquad a=1,...,16\, ,
}
where the entries $\Pit^\La$ with $\La>2$ are identified with the vector multiplet scalars $t_a=(\SS,U,C_b)$.
In the orientifold limit \ofi, these scalars correspond\foot{Up to a subtlety in the quantization
condition \FF.} to the type II dilaton $\SS$, the complex structure $U$ 
of $T^2/\IZ_2$ and the positions $C_a$ of 16 D7-branes. The superpotential in the scalars $t_a$ is then
\eqn\supiii{\eqalign{
W_F\ &=\ \ p_1+p_2\, (-U  \SS+\fc{1}{2}\, C_aC_a )+p_3\, \SS  +p_4\,  U 
+\sum_{a=1}^{16}p_{4+a} C_a\cr
&\sim \ \ \int \Omega \wedge G_3,\qquad {\rm for}\ \ \ C_a=0,\ \forall a.
}}
where $p_\La=\Pi^I \g_{I\La}$. As indicated, upon setting $C_a=0$, the expression reduces 
to the orientifold potential \spst\ for \ofi. In particular the 
four fluxes $\cx F_\La$, $\La<5$ are identified with the bulk 3-form fluxes $F_3,H_3$ with 
one leg along one of the two 1-cycles of $T^2/\IZ_2$ in the orientifold limit. The remaining 16 fluxes
correspond to 2-form fluxes on the 16 D7-branes of the orientifold. Note that the potential from the bulk fluxes,
corresponding to the first four terms, contains a quadratic term in  $C_a$ proportional to $p_2$ that is not 
reproduced by \spst. It gives a mass to all D7-brane moduli \GKTT\FF.\foot{See also the
computations of soft-terms on the world-volume in \LST\IBW\IBWii\LJ.}

\subsec{Inclusion of instantons}
The potential \supiii\ can be rewritten in terms of the
effective supergravity form,
$$
W_F=P_\La^{hol}X_0^\La,
$$
using the identifications
\eqn\fres{
X_0^\La=\Pit^\La,\qquad P^{hol}_\La = \Pi^I\, M_{IJ}\, k^J_{\ \La},\qquad k^I_{\ \La}=M^{IJ}\g_{J\La},
}
where indices can be raised and lowered with the help of the 
intersection matrix $M^{IJ}=\int \eta^I\wedge\eta^J$ and its inverse. The 
constants $k^I_{\ \La}$ are the Killing vectors that describe the charge of 
the hyper multiplet scalars
under the $\La$-th $U(1)$ gauge symmetry; see \thebible\ for details on the role of the 
$k^I_{\ \La}$ in supergravity. The above relations replace the eqs. \clrel\ in the orientifold case.

As mentioned previously, in the closed string case the period integrals 
compute the exact symplectic sections $(X^\La,F_\La)$ and the geometric result is already the
final answer. There is no similar argument in the case with background branes and one expects
further corrections. In fact the candidate 4d superpotential $W_F$ depends very weakly on the
extra compactification; in particular it does not depend on the
volume modulus $\TT$ of the lower K3 $\kb$. 
For the constant Killing prepotential $P^{hol}=1$, it has the strange property that it 
is in fact completely independent
of the compactification. 

The reason is that the {\it leading} effect of the compactification
is exponential in the K3 volume $\TT$ 
and is not reproduced by the naive dimensional reduction.
Exponential terms in $\TT$ indicate a non-perturbative origin from the point of the type II string 
and such terms may arise from Euclidean D3-brane instantons wrapped on $\kb$ \Witsp. 

In the next two sections we use a string duality to study the quantum corrections from D3 instantons
and we find that these corrections depending on $\TT$  can be written in the factorized form
\eqn\fullsp{\eqalign{
W_{eff}&=P_\La^{hol}X^\La=P_\La^{hol}(\, X_0^\La+X_{qu}^\La\, )=
\phantom{\pmatrix{1\cr 1}}\cr
&=W_F(\g,q^i,t_a)+\Wi(\g,q^i,t_a)\, .
}}
Here $W_F$ is the quartic polynomial \supiii\ and the instanton correction $\Wi$ has the general form
\eqn\defWi{
\Wi= \Pi^I(q^i)\, \g_{I\,2}\, \ffi(q_a),\qquad \ffi(q_a)=\sum_k \qtt^k f_k(q'_a),
}
where\foot{The K3 volume $\TT$ is also a scalar in a  4d vector multiplet; we use
here and in the following a prime to denote the vector scalars $t_a'$ excluding $\TT$.}
$$
\qtt=\exp(2\pi i \TT),\qquad q'_a=\exp(2\pi i t'_a).
$$ 
Specifically a term $\sim \qtt^k$ in \defWi\ corresponds to a D3-brane instanton
of charge $k$. 

The non-zero, and as we will see highly non-trivial D3-instanton sum $\Wi$ in the superpotential 
is somewhat surprising because there are no divisors in the simple geometry \ktk\ that satisfy
the necessary condition of ref.\Witsp, $\chi=1$. However, as already pointed out in \GKTT,  
the discussion in \Witsp\ does not take into account the effect of the flux on the zero-mode counting. 
It was also suggested there how physical expectations on the dynamics of 
$\cx N=1$ SYM predict a non-zero contribution.

The divisors that support the D3-instanton corrections, as computed in the following
sections, have been listed in sect.~2. The details of these instantons are somewhat 
different from the situation described in \GKTT, as the divisors have a different topology, there is 
no non-Abelian phase and the instantons carry integer charges. However there is no contradiction at all;
the instantons we find are closely related to the instantons discussed in \GKTT\ by the deformation of 
$\cx N=2$ theories considered in \SW.

Let us briefly discuss two non-trivial features of the instanton corrections anticipated by the ansatz \defWi.
First note that only {\it one} flux component, $\La=2$ in our conventions, induces instanton
corrections to the superpotential. This has a simple physical meaning. As discussed before,
the 3d
instanton has a magnetic charge under a particular $U(1)$ gauge group. The flux 
that lifts the zero modes on this instanton is, not surprisingly, a flux for the gauge field
in the same $U(1)$. In the orientifold limit, the gauge flux translates to the 
3-form flux component 
$$
H_3=dx\wedge \al_x,
$$
with $\al_x$ an integral 2-form on $\kb$. 

The result that the superpotential is proportional to a 
single flux component is seemingly in contrast to the results from
the world-volume computations in \TTii\Saulina\KPT. One reason is that the 
field basis defined by  the world-volume action is 
unphysical in the sense that it is not simply related 
to physical observables in the non-perturbative theory. 
A more meaningful description of the physical instanton 
corrections is in terms of non-perturbative coordinates
on the moduli space that are related to physical, measurable quantities in 
a simple way.
A natural way to define such coordinates is 
in terms of the non-perturbative BPS tensions of physical domain 
walls, and this is the definition used in this paper; see also 
sect.~6 for a further discussion. 

Secondly, the ansatz \defWi\ constrains already to a large extent the moduli dependence 
of the complex structure moduli. Note in particular,  that the corrections from 
the D3-brane instantons wrapped on $\kb$ preserve the factorized form of the superpotential,
as they enter only in the symplectic section $X^\La$.  This factorization will play an important
role when we define a generalized flux potential for the type II side in sect.~6; 
a more detailed discussion of the moduli dependence will be given in sect.~5.

\subsec{$D$-term potential}
As already mentioned, the same flux \Fflux\ generates also a $D$-term potential \FF\LJ.
This potential restricts K\"ahler moduli on the lower K3 $\kb$:
\eqn\defvd{\eqalign{
V_D&=\fc{1}{2}e^{K_H}\   P^3_\La \, N^{\La\Si}\, P^3_\Si = \fc{1}{2}\,D_\La N^{\La\Si} D_\Si,
}}
where $N^{\La\Si}$ is the inverse coupling matrix of the gauge fields \thebible\ 
and the real
potentials entering the $D$-term $D_\La=e^{K_H/2}P^3_\La$  are 
\eqn\Dterm{
P^3_\La = \Pi^{3,I}(q^i)\, \g_{I\La}=\fc{1}{\sqrt{V}}\ \int j\wedge \cx F_\La,
}
where $j$ is the K\"ahler form on $\kb$, $V=\fc{1}{2}\int j\wedge j$  and $\cx F_\La$ the flux \Fflux.
The second expression in \defvd,  is the standard form of the $D$-term in $\cx N=1$ supergravity.

The period vector $\Pi^{3,I}$ is defined in \FF\ as the integral of a self-dual 2-form $\om^3$,
with norm one in a finite volume K3. The content of the $D$-term equation $D_\La\sim \Pi^{3,I}\g_{I\La}=0$ 
is that the 2-form $\eta^I\g_{I\La}$ is orthogonal to $\om^3$. Each $D$-term which corresponds to an 
anti-selfdual 2-form restricts one K\"ahler modulus of $\kb$ at finite overall volume. This leads to 
a quite effective stabilization of K\"ahler moduli.\foot{For a similar statement for systems with D9/D5
instead of D7/D3 branes, see \AM.}
However a $D$-term proportional to $\om^3$ can not be solved at finite volume. A $D$-term in this directions
leads to a decompactification (zero coupling) limit in the absence of other effects. 

\subsec{Comparison with the (A)dS ansatz of KKLT}
Before computing the explicit instanton corrections, 
let us compare the general form of the potential \fullsp\ to the KKLT ansatz \KKLT. 

Firstly there is the quartic polynomial superpotential
$W_F$ in \supiii, that depends for generic choice of fluxes on all 
complex structure moduli of the geometry and the type II dilaton $\SS$. In the orientifold
limit it reduces to the superpotential \spst\ of \GKP\KKLT. The polynomial superpotential  
$W_F$ does not depend on the volume modulus $\TT$ of the lower K3, $\kb$. 

The second ingredient is the non-perturbative correction $\Wi$ which introduces a dependence
on the volume modulus $\TT$. It combines with the classical part $W_F$ to the exact section 
that determines the effective superpotential. The first term $\sim \qtt$ in the expansion $\Wi$ should be 
compared with the non-perturbative term in \KKLT,
$$
\Wi|_{k=1}=Ae^{2\pi i \TT},\ \ \qquad A=\Pi^I(q^i)\, \g_{I\,*}\, f_1(q'_a),
$$
where $*$ denotes the single flux component that contributes to the superpotential, $\La=2$ in the 
conventions of \pvkv.
The coefficient $A$ it is linear in the flux $\g_{I*}$, in agreement with the expectation that the D3 instanton
may contribute to the superpotential only if there is a non-zero flux that lifts  zero modes. 
The dependence on the moduli $q^i$ and $q_a'$ arises from the one-loop determinant \Witsp. 

The third ingredient in the ansatz of KKLT is a small positive energy contribution
that lifts AdS vacua to dS vacua. In \KKLT\ this contribution comes
from anti-D3-branes sitting in a sufficiently warped throat, but there are 
also several other possibilities \Silv\BKQ\BaBe. In the present compactification such 
a contribution is also included already from the beginning in the form of the 
$D$-term potential \Dterm, realizing a variant of the variant of \BKQ.\foot{The 
moduli dependence of the potential $V_D$ in eq.\defvd\ is different from that in \BKQ, since the 
axions which receive a non-zero $U(1)$ charge in the flux background are not
in a multiplet with the volume modulus $\TT_2$. The superpartners of the axions 
are the K\"ahler moduli of $\kb$ at fixed volume, and the 8d real scalar that
measures the volume of the base $B_V$ of the elliptic K3 $\kf$.}

To summarize, just switching on F-theory flux in the simple geometry \ktk\ is sufficient 
to generate all three ingredients of the KKLT ansatz, namely the polynomial potential for the
complex structure moduli $W_F$, an (infinite, computable) D3 instanton sum $\Wi$ and a
positive energy contribution $V_D$ that may lift the vacuum energy to positive values. 
All three pieces arise from the same 2-form flux described by the single matrix $\g_{I\La}$
and fall natural into place. We will now turn to a computation of the instanton corrected
superpotential \fullsp\ by an open-closed string duality.

\newsec{Open-closed String duality}
\subsec{Duality}
The compactification of F-theory on \ktk, or its orientifold limit, the type IIB on \ofi,
leads to a $\cx N=2$ supersymmetric theory in four dimensions. This is the same supersymmetry
as that of a type II closed string compactification on a Calabi--Yau 3-fold $Z$, without any background
branes. By the general properties of the $\cx N=2$ supersymmetric moduli space, one expects that 
the two compactifications can be dual to each other. For a trivial 
gauge background, one can rely on known F-theory/heterotic dualities to infer such a duality \SVW.
More general cases of such $\cx N=2$ dual pairs will be described below. 

Let us briefly recall the argument of \SVW.
The basic duality is the one between 
F-theory on an elliptically fibered K3 $\kf$ 
and the heterotic string on $T^2$ in 8 dimensions \vafaf. Compactify on a 
further K3 $\kb$ with trivial gauge fields. There is a tadpole in four dimensions canceled
by 24 D3-branes on the F-theory side and 24 5-branes wrapped on $T^2$ on the heterotic side. 
On the other hand, the heterotic string on $\kb\, (\times T^2)$ is dual to F-theory on $Z\, (\times T^2)$,
with $Z$ an elliptically fibered 3-fold \MV. Including the $T^2$ factor, 
this is also the same as type IIA on $Z$
or the type IIB string compactified on the mirror manifold $\Zh$. The hodge numbers of $Z$ are $h^{11}=43=h^{21}$.
On the F-theory side there are 43 vector multiplets in four dimensions,
with 19 gauge fields from D7-branes and
24 from D3-branes. There are also 44 hyper multiplets, 20
from the K3 with background fields and 24 from the positions of the D3-branes on K3. 

The above compactification with a trivial gauge background can be connected to non-trivial gauge backgrounds
by moving $k$ D3-branes onto a 7-brane stack with non-Abelian gauge group $G$ 
and moving onto the Higgs branch in the gauge theory. This corresponds to giving vevs to the 
hyper multiplets from 3-7 strings and represents a finite size $G$ instanton of degree $k$ \Douglasbrane.
Such transitions have been studied in \BS, together with more exotic Higgs branches.

The various branches with different number of D3-branes, 
non-Abelian gauge symmetries, possibly broken by finite size instantons,
lead to many components of the $\cx N=2$ moduli space that differ in the massless and massive spectra.
They are connected by transitions that can be locally understood as (un-)Higgsing. 
The F-theory compactification with a generic gauge background on \ktk\ and at generic moduli 
has a well-defined dual closed string compactification, with different components 
of the moduli space described by different \CY\ manifolds $Z$. 

On the other hand clearly not any $Z$ will correspond to a perturbative F-theory compactification on \ktk.
As discussed in the previous section one characteristic feature is the peculiar dependence
on the modulus $\TT$ of the K3 volume. As discussed in sect.~4.3 below, this requires
that the dual manifold $Z$ is a K3 fibration, which also follows from heterotic/type IIA duality~\KV.\foot{There
are also F-theory compactifications associated with non K3 fibrations, but those are strongly coupled
and do not have a simple interpretation as a compactification on \ktk.}

\subsec{F-theory flux and generalized type II compactifications}
Having a dual $\cx N=2$ pair one may add deformations to break supersymmetry. Since the parent theories
are supposed to be dual, turning on quantized background fields, such as flux or metric deformations,
should also match on the two sides. However the meaning of a certain deformation may be 
very different in the two different compactifications. To compare the deformed backgrounds we will
describe both theories within a common framework, namely their underlying effective 4d 
$\cx N=2$ supergravity theory. The effect of the background flux can be described by certain gaugings in 
the effective supergravity \PS. The effective supergravity then allows to switch between 
the descriptions in the two dual theories.

The mapping from the F-theory fluxes to gaugings in supergravity has been (partially) worked
out in \FF, and is summarized in \fres. We will use this information to define (a subset of) those gaugings
in supergravity that can be represented in the string theory compactification.
In fact the gaugings of the type II string realized by mapping the F-theory fluxes to 
data on the manifold $Z$ leads to substantial generalizations of type II compactifications.
The gaugings in the effective supergravity relevant for such compactifications have been described recently in \FerH.
We will discuss the type II side of these compactifications further in sect.~6, where we write down
a generalization of the GVW superpotential \spst\ for these compactifications. Note that it follows from 
F-theory that these compactifications still have supersymmetric branches, {\it c.f.} \TriTri\FF. 

On the  other hand a systematic computation of the instanton terms is, at present, only possible in the closed string
compactification on the Calabi--Yau 3-fold $Z$. As usual,  combining the two dual pictures one gets 
valuable information for both theories.

\subsec{Effective supergravity for the closed string dual}
The computation of the instanton corrected superpotential will be performed in the 
dual closed string compactification of type IIA on a CY\ manifold $Z$ and 
the type IIB compactified on the mirror manifold  $\Zh$.
In the following we give a brief description of the relevant aspects of 
the effective supergravity for $Z$ and its relation to the F-theory compactification.

In the type IIA compactification on the \CY\ manifold $Z$,
the scalars $t_a$ in the vector multiplets represent 
$h^{11}$ complexified K\"ahler parameter. The 
holomorphic $\cx N=2$ prepotential for these K\"ahler moduli is
\eqn\cypp{
\cx F=\fc{1}{6}C_{abc}t_at_bt_c-\fc{1}{2}B_{ab}t_at_b-A_at_a-i\chi\fc{\zeta(3)}{2(2\pi)^3}+ f(q_a),
}
where $\chi$ is the Euler number of $Z$ and 
the coefficients of the first three terms are related to topological data of $Z$ as
$$
C_{abc}=\int_Z J_a\wedge J_b\wedge J_c,\ \ B_{ab}= \fc{1}{2}\int_Z J_b \wedge i_*c_1(E_a),
\ \ A_{a}= \fc{1}{24}\int_Z J_a \wedge c_2(TZ),
$$
where $\{J_a\}$ is a basis for $H^{1,1}(Z)$, $E_a$ is the divisor dual to $J_a$ and 
$B_{ab},A_a$ are defined mod $\IZ$ \Rtopdata.
Finally, the term $f(q_a)$ contains the contribution from world-sheet
instantons that depends only on the exponentials $q_a=\exp(2\pi i t_a)$. A subset of these
terms will be matched to the superpotential in F-theory from D3-instantons below.

The holomorphic prepotential $\cx F$ for the \CY\ manifold $Z$ defines a standard symplectic 
section $(X^\La,F_\La)$ of the effective supergravity  as 
\eqn\stansec{
X^\La=\pmatrix{1\cr t_a},\qquad F_\La =\pmatrix{2\cx F-\sum_a \cx F_at_a\cr\cx F_a}
}
where $\cx F_a=\p_{t_a} \cx F$. 

Recall that the F-theory compactification on \ktk\ depends in a very 
special way on the volume modulus $\TT$ of the lower K3, $\kb$.
Specifically, the upper half $X^\La$ of the symplectic section, obtained in eqs.\pvkv\fres\ from F-theory, 
does not depend on $\TT$. This is compatible with the CY expression only if 
there is a symplectic rotation of the section \stansec\ such that the upper half $X^\La$ of
the new section is independent of one modulus $t_{\hx a }$ that can be identified with $\TT$. 
It is not difficult to see that this implies that
the component $F_{\hx a }$ of the section 
dual to $t_{\hx a}$ is independent of $t_{\hx a}$, which means that
the prepotential $\cx F$ is linear in $t_{\hx a}$. 

This condition has been well-studied in the context of heterotic/type II duality and 
it implies that $Z$ is a K3 fibration \KLM\AL. The {\it classical} supergravity 
corresponding to these class of manifolds is based on a special
K\"ahler manifold for the vector multiplets of the type \CF\ADFL
\eqn\msugra{
\cx M_V=\big(\fc{SU(1,1)}{U(1)}\big)_\TT \times \big(\fc{SO(2,n)}{SO(2)\times SO(n)}\big)_{t'_a}.}
In the F-theory compactification on \ktk, the second factor is to be identified with the 
8d moduli in eq.\pvkv\ 
from the complex structure of the elliptically fibered K3 $\kf$ with $n=18$.\foot{See \vafaf\aspe\FF\
for details on the K3 moduli.} 
On the other hand, the first factor 
will correspond to the K3 volume $\TT$. In the CY manifold this modulus describes
the complexified volume of the base $B_Z$ of the K3 fibration $Z\to B_Z$,
$$
\rho=\int_{B_Z} B+iJ,
$$
with $B$ the NS 2-form..
The prepotential for the factorized space
\msugra\ is linear in $\TT$ and the first derivative of the CY prepotential \cypp\ of a 
K3 fibration looks like 
\eqn\Frho{
\cx F_{\TT}= (\, \SS\, U -\fc{1}{2}\, C_a^2\, )-A_{\TT}+\ffi(q_a),\qquad \ffi(q_a)=\p_{\TT} f(q_a),
}
where it follows from  the CY geometry that $A_{\TT}=1$ \AL.
After a symplectic transformation $(\TT,\cx F_{\TT})\, \to \, (\cx F_{\TT},-\TT)$, the new section
$(X^\La,F_\La)$ is
\eqn\newbas{
X^\La=\pmatrix{1\cr (\, \SS\, U -\fc{1}{2}\, C_a^2\, )-1+\ffi(q_a)\cr\SS\cr U\cr C_a}=
\Pit^\La+\cx \delta^\La_2 \ffi(q_a).
}
As indicated, the classical part of the section $X^\La$ for the \CY\ $Z$ 
coincides with the period integrals $\Pit^\La$ 
of the upper K3 $\kf$, eq.\pvkv, in the dual F-theory compactification.\foot{Up to a 
perturbative symplectic rotation on $X^\La$.}
Moreover we have already anticipated the map from the remaining moduli $t_a'$
to the F-theory moduli in \pvkv. In particular, $\TT$ will be identified with the K3 volume modulus of 
the same name in F-theory, and  in the orientifold limit
the scalars in the bracket map under the duality to
the type II dilaton $\SS$, the complex structure
$U$ of $T^2/\IZ_2$ and the positions $C_a$ of the 16 7-branes. 

The above discussion is oversimplified as we have not considered further CY moduli $t_a$ 
that are dual to D3-brane positions or to non-perturbative states. However these moduli
can be added easily at a later stage and we will do so to keep the discussion simple
at this point; see also Table~1 below.

The basis change between the CY basis and the symplectic basis \newbas\ is also familiar in the 
context of heterotic string dual in \CF. Note that for F-theory on \ktk, 
the electric section $X^\La$ in \newbas\ is just the standard section for the electric gauge fields
on the 7-branes, whereas the dual \CY\ section \stansec\ mixes 1- and 5-form
gauge fields from the 7-branes.

The term $\ffi(q_a)$ in \newbas\ describes the D3-instanton corrections defined in \defWi,
as computed by the closed string dual on $Z$. Note that its leading term is $\cx O(\qtt^1)$ and
therefore this correction vanishes as $\TT \to i \infty$, in agreement with the fact, that
the only correction to the superpotential in the vector scalars comes from D3 instantons
wrapping $\kb$. 
Moreover only the period of the $U(1)$ vector multiplet that couples to the monopole 
is corrected by the D3-instantons, justifying the ansatz in \defWi.

On the other hand, the second half $F_\La$ of the new symplectic section is given by 
$$
F_\La = \TT\, \tx M_{\La\Si}\Pit^\Si + ...
$$
where $\tx M_{\La\Si}$ is the inverse of the intersection matrix on $H^2(X,\kf)$.
It receives exponential corrections that contain a non-zero term at order $\cx O(\qtt^0)$ from 
$$
f_0(q'_a)=f(\qtt=0,q_{\SS},q_U,q_{C_a})
$$
and its derivatives. 
Although the corrections in $f_0(q_a)$ do not contribute to the superpotential
they enter other quantities, namely the gauge couplings and the K\"ahler potential
\eqn\kp{
K_V=-\ln (-i\, Y),\qquad Y=2\, (\cx F-\bb{\cx F})-(t_a-\bb t_a)\, (\cx F_a + \bb{\cx F_a})\ .
}
These are the contributions of the instantons to the amplitudes in the parent theory without flux
discussed in sect.~2. This has the effect of changing the values at which the $t_{a'}$ are 
stabilized even in the limit  $\TT \to i\infty$. Moreover, even the classical part of the 
K\"ahler potential has a more complicated dependence on the moduli $t_a$ than in the 
KKLT model. This leads to an important enhancement of the instanton contribution, 
as discussed in sect~7.

\subsubsec{Space-filling D3-branes}
It is easy to generalize the above discussion to include $k$ space-filling D3-branes. The simplest
way to see what they correspond to in the \CY\ $Z$ is to use the heterotic dual on $K3\times T^2$, 
where the $k$ D3-branes map to $k$ 5-branes wrapped on $T^2$.
Decompactifying the $T^2$, one obtains a heterotic string on K3 with $n_T=k$ extra tensor multiplets
from the world-volume fields of the 5-branes. From \MV\ we know that extra tensors arise from 
$h^{11}(B_2)$, where $B_2$ is the base of the elliptic fibration $Z\to B_2$. Compactifying back 
to 4d on $T^2$ we see that the $k$ D3-branes in 4d are described by $k$ elements of $H^2(B_2)$
in the type IIA theory on $Z$. 

Moreover, as mentioned already, space-filling D3-branes can be
obtained by a process of un-Higgsing in the 7-brane gauge theory, where
an instanton on the 7-brane shrinks to zero size and moves onto the Coulomb branch
of the D3-brane world volume theory \Douglasbrane.
Geometrically this process of creating a D3-brane 
corresponds to a blow up of the base $B_2$ and by iteration, 
the $k$ elements in $H^2(B_2)$  correspond to $k$ blow ups of a minimal base $B_2$ describing the
case with zero D3-branes. The more general statement for theories without a 
6d limit is that D3-branes map to blow ups of the K3-fibration $Z\to B_Z$.\foot{One way
to show this is to start from a K3 and elliptic fibration with an 6d limit, and perform an
extremal transition of the type \BKKM\ that leaves only a K3 fibration.}
The mapping of the moduli is sketched in the table below.\foot{We do not consider
compactifications with non-perturbative gauge fields $C'$, which arise from 
reducible elliptic fibers that are are not contained in the K3 fiber; see
\MV\ for a discussion.}

\vbox{$$
\vbox{\offinterlineskip\tabskip=0pt\halign{\strut
~$#$~\hfil&~$#$~\hfil\hskip10pt&~#~\hfil\cr
\cr
\#& {\rm moduli} &F-theory on \ktk\cr
\noalign{\hrule}
3\ {\rm generic}&\TT;\ \SS,\ U&K3 volume; 2 D7-branes \cr
h^{1,1}(Z)-h^{1,1}(B)-1\qquad &C,\ (C')&D7-branes\ (non-pert.)\cr
h^{1,1}(B)-2&B_\al&D3-branes\cr
}}
$$
\noindent{\ninepoint  \centerline{{\bf Tab.~1:} Mapping between moduli between $Z$ and F-theory on \ktk.}}\vskip10pt
}
\noi
Let $\BB_\al$ denote the $k$ extra vector multiplets from the $k$ space filling D3-branes. 
We will discuss the map between the D3-branes moduli and K\"ahler moduli in great detail
in sect.~5, when we compute the classical supergravity generalizing \msugra, 
and the instanton corrected superpotential. 
However let us note here that if a 6d limit exists, the couplings of these extra vector
multiplets to the universal fields $\TT,\SS,U,C_a$ are strongly constrained \AFM.
See sect.~2 of \LSTY\ for a nice discussion of the properties of these couplings in the 
context of the dual heterotic string compactified on $T^2\times K3$ and a list of further references.

\subsec{Instantons and fluxes}
As discussed in the previous section, the volume modulus $\TT$ of the K3 $\kb$ maps to the volume 
of the base $B_Z$ of the K3 fibration of the manifold $Z$ under duality. Comparing BPS actions, 
it follows that a D3-instanton wrapped on $\kb$ maps to a world-sheet instanton wrapped on $B_Z$.

To gain further understanding of the nature of this instanton it is helpful to consider the
heterotic dual \Witsp. Since the F-theory and the heterotic string are already dual in 8d,
the branes that represent the instantons must wrap the same cycles
on the lower K3 $\kb$ in the dual theories .

D3-branes wrapping a divisor of the type $C)$ wrap a curve $C_1$ in $\kb$ and correspond to 
heterotic world-sheet instantons, by the above map or a comparison of BPS actions.
These instantons correct the metric of the hyper multiplets. 
We will not have to say much about these corrections, 
although they can be also computed in some cases, as 
discussed briefly in sect.~6.

A D3-brane instanton that wraps a divisor of type $A)$ wraps all of $\kb$ and 
the dual object is the heterotic 5-brane wrapped on K3$\times T^2$ \Witsp.
Using heterotic/IIA duality one recovers the above identification of the D3-instantons 
with world-sheet instantons on $B_Z$ in the following way. The heterotic 5-brane  wrapped on $T^4$ is the dual 
of the IIA string on K3 in 6d \HS. If $\kb$ is elliptically fibered, 
the heterotic compactification can be viewed as a compactification on $T^4$ fibered over 
the $\IP^1$ base of $\kb$, which is the same as the base $B_Z$ of $Z$, thus completing the
duality circle.

On the other hand the above argument indicates, that the instanton obtained by wrapping the D3 brane on $\kb$
has an intricate structure. The heterotic 5-brane wrapped on K3 gives a string in 6d; the nature of the
string depends on the details of the background, and we defer a discussion of it to the next section
where we consider specific compactifications. 
Compactifying this string on a circle yields an infinite number of massive 5d BPS particles. 
The 4d instantons to the superpotential arise from world-lines of these particles
wrapped along an extra $S^1$. Thus the basic instanton in the superpotential comes actually with
an infinite sum of dressings by the massive modes; this expectation
is confirmed by the computation in the next section.

\vbox{$$
\vbox{\offinterlineskip\tabskip=0pt\halign{\strut
~#~\hfil\hskip15pt&~#~\hfil\hskip15pt&~#~\hfil\hskip15pt&~#~\hfil\cr
\cr
F/\ktk& IIA/$Z$ &Het/$T^2\times$K3&IIB/$\hx Z$\cr
\noalign{\hrule}
D3 on $A)$&F1/$B_Z$&NS5/$T^2\times$ K3 & \cr
D3 on $C)$&        &F1/$C_1\subset$ K3 &F1/$C_1\subset \hx Z,...$\cr\cr
}}
$$
\leftskip45pt
{\ninepoint {\bf Tab.~2:} Basic mapping of instantons  in the duals.
The cases $A)$, $C)$ \br refer to the type of
divisors discussed in sect.~2.}
\vskip10pt
}

\subsubsec{Flux}
The flux \Fflux\ arises from a 4-form flux in M-theory of the form 
$$
G_4=\cx F_\La \etat^\La+\cx F_0\eta_B,
$$ 
where $\etat^\La$ is an element of $H^2(\kf)$ orthogonal to the class of the elliptic fiber $E$ 
and the class of $B$ of the section of the elliptic fibration $\kf \to B_V$. 
Simply speaking, the fluxes in the directions $\etat^\La$ have
one leg on the fiber $E$ and one leg on the $B$. A tadpole from this flux can therefore only arise
for the divisors of type $B)$. On the other hand there is one extra flux $\cx F_0$ dual to
the base $\IP^1$ of $\kf$ \Witsp\FF. This flux may induce tadpoles and obstruct the 
instantons wrapped on the divisors of type $C)$ that correct the hyper multiplet metric.
Finally, only the flux in the single direction ($\La=2$) of the monopole $U(1)$ generates the
D3-instanton sum.

\subsubsec{$\cx N=1$ generalizations}
It is natural to generalize the above duality to F-theory compactifications
with $\cx N=1$ supersymmetry in the absence of flux.
It is interesting to note that the map from D3-branes to world-sheet instantons applies
also to this more general case, at least for a certain class of D3-branes. 
One can use an adiabatic argument, or one may consider orbifolding
using the Voisin-Borcea construction. E.g. one may  start with a configuration,
where the F-theory on \ktk, the heterotic string on $T^2\times K3$ and the manifold $Z$ are invariant under
a common $Z_2$ involution, which includes an action on the gauge fields.
Modding by the involution one obtains a duality between F-theory on a 
\CY\ 4-fold $X_4\simeq K3\times K3/\IZ_2$, heterotic string on a \CY\ 3-fold $X_3\simeq T^2\times K3/\IZ_2$,
a type IIA orientifold on $Z$ and F-theory on a \CY\ 4-fold $X'_4\simeq T^2\times \Zh/\IZ_2$. 

\newsec{Duality at work}
We compute now the D3-brane corrections to the superpotential
in the dual closed string compactification of the type IIA string on a 
\CY\ manifold $Z$. Specifically we compute the exact sections $X^\La$
in \fullsp\ and relate them to the dual D3-instantons in the F-theory compactification.

As explained before, the D3-instanton correction depends in general on the
other moduli of the compactification of F-theory on \ktk\ through the 1-loop determinant. In particular
the exact section $X^\La(t_a)$ depends on all scalars in the vector moduli $t_a$. 
To write the flux potential in all these moduli leads inevitably to complicated and 
not very illuminating expressions. One can fix most of the vector moduli $t_a$
already in the $\cx N=2$ supersymmetric compactification without flux,
by switching on finite size instanton bundles on K3 \MV\ and we will do so. 
Note that the expression \supiii\ does not lead to a potential for these gauge background
as it depends only on the Abelian part of the gauge fields.

As discussed above, a \CY\ manifold $Z$ that describes a type II compactification 
dual to the F-theory compactifications on \ktk\ must have a field $\TT$
that appears linearly in the prepotential and this requires that 
$Z$ is a K3 fibration. It is also useful to require the existence of a
6d limit, corresponding  to F-theory on $Z$ and a perturbative heterotic dual on $K3$.
The minimal number of  moduli $t_a$ for an elliptic and K3 fibered \CY\ manifold $Z$ is three,
and these correspond under the duality to two 7-brane moduli $\SS$ and $U$, and in addition the volume modulus
$\TT$. Space-filling D3-branes  can be added straightforwardly, as will be done in sect.~5.2.

\subsec{2 D7-branes}
The minimal case with two independent D7-branes is described by elliptic fibrations $Z_n$ over Hirzebruch surfaces 
$F_n$, $n=0,1,2$. The definition of these manifolds as toric hypersurfaces, and the map from the 
three K\"ahler moduli $t_a$ to the F-theory moduli $(\SS,U,\TT)$ is summarized in app.~A.1.

As already written in \defWi, the D3-instanton sum is of the form 
\eqn\defWi{
\Pi^I(q^i) \g_{I*} \sum Q^k\ f_k(q'_a).
}
Note that the dependence on the complex moduli $q^i$ of the lower K3 $\kb$, wrapped by the D3-brane, is universal. 
On the other hand, the dependence on the complex moduli $t_a$ of the upper K3 $\kf$ depends on the details
of the instanton. 
In the language of the M-theory 5-brane, the coefficient functions $f_k(q'_a)$ encode the sum over
field configurations on the  world-volume depending on $H^3(D)$ \Witfb. For the divisor $D=T^2\times \kb$, 
an element of $H^3(D)$ is of the form $\eta^I \wedge \al_k$,
where $\eta^I\in H^2(\kb)$ and $\al_k$, $k=1,2$  are the 1-forms on the elliptic fiber. 

Let us briefly sketch how the dependence on the moduli $t_a'$ arises in the language of F-theory.
The D3-instanton wrapped on $\kb$ has a moduli space, which is parametrized by the position 
$z$ on the base $\IP^1$ of the elliptic fibration  $\kf \to B_V$. The instanton exponential
for the D3-brane wrapped on $\kb$ is 
\eqn\iwei{
e^{-Vol(\kb)/\la(z,t'_a)\ ,}
}
where $\la(z,t_a')$ is the local value of the type IIB dilaton. In the F-theory compactification 
$\la$ is not a constant, but it varies with the position $z$ on $B_V$. 
The value of the coupling constant $\la(z,t'_a)$ at a position $z$
can be computed as the effective coupling $\tau$ of the $\cx N=2$ supersymmetric gauge theory
living on a D3-brane probe at $z$ \Sen\BDS. The theory on the D3-brane probe depends on the location of 
the 7-branes, e.g. through the states associated with a string stretched between the D3-brane and the 7-brane. 
The contribution of an instanton of fixed charge $k$ is the integral over the positions on $B_V$,
and it receives a dependence on $t_a'$ through the dependence of the local value of $\la(z,t_a')$ 
on the 7-brane positions. Note that there is the standard logarithmic singularity in the 
coupling $\tau_{eff}$ when $z$ is near a 7-brane. This suppresses the regions, where the 
D3-instanton is near the 7-branes and leads to a rather soft dependence of the coupling on 
$t_a'$, for generic values of these scalars.

Although the full dependence in \defWi\ can be computed from the type II string, let us first
switch off the effect of the 7-brane moduli $t_a'$ by making them very large. In this limit, 
$f_k$ becomes a constant, in agreement with the fact 
that the coupling $\la$ is approximately constant
and moreover the patches where the D3-instanton is near the 7-branes decouple.

As discussed in sect.~2, the contribution in this limit $q'_a\to 0$, is expected to coincide with
the first integral of the gauge coupling of the monopole multiplet
\eqn\ismodi{
\ffi(\qtt)={b_n\over (2\pi i)^2}\sum_{k=1}^\infty \fc{\qtt^k}{k^2},
}
where $b_n$ is the beta function of the monopole multiplet. The value of $b_n$ can be determined
by studying the gauge theory in the type IIA string when the volume of the section $B_Z$ of the
K3 fibered manifold becomes small. 
The nature of this gauge theory depends on the normal bundle
of $B_Z$; for the cases $n=0,1,2$, one obtains a pure $SU(2)$ gauge theory, a theory of $U(1)$
with one electron and a $SU(2)$ theory with an adjoint matter multiplet, respectively \Witpt\refsin,
leading to the values 
\eqn\beta{
b_n=(-1)^{n}\, (n-2).
}
This is confirmed by a computation of the instanton sum $\ffi$ from 
the dual type IIA string theory compactified on $Z_n$. The sum  \ismodi\ arises from a single 
non-zero Gromov-Witten invariant $$N_{0,0,1}=b_n$$ and describes 
a world-sheet instanton wrapped on $B_Z$, together with its multi-coverings.\foot{The 
number $N_{\vec n}$ is defined as the coefficient in the 3-point function,
$$
\p_a\p_b\p_c \cx F=C_{abc}+\sum_{\vec n}\fc{n_a n_b n_c\, N_{\vec n}\prod q_i^{n_i}}{1-\prod q_i^{n_i}}
$$
with $\vec n=(n_1,...,n_{h^{11}})$ and $n_k$ non-negative integers. See \mirrlec\ 
background material and references.}

This confirms that not just the singly wrapped D3-brane instanton contributes
to the superpotential \fullsp\ of 
the dual F-theory, but also all multi-wrappings contribute with an extra weight factor $1/k^2$.

The instanton sum depends further on the $t_a'$ for finite values. These corrections have
a fascinating interpretation in terms of the non-trivial world-volume theory on the 
5-brane in M-theory, or on the D3-brane in F-theory. The latter is a twisted 
$\cx N=4$ supersymmetric gauge theory that has been studied in \VWx\MNVW.

In fact,  all (but one) of the above instanton corrections, as well as those in the $\cx N=1$ examples
of F-theory on 4-folds studied so far \Witsp\PMff\grassi\DDF\DK, 
can be interpreted as world-sheet instantons of a certain non-critical string \PMff.
One can see this string in any of the dual formulations;
let us take the type IIA compactification on $Z$ for concreteness. This compactification 
is related to F-theory on $Z$ in 6d, where the non-critical string in question is made
by wrapping the D3-brane of type IIB on the section $B_Z$. The nature of the
string depends on the manifold $Z_n$. There is a nice and simple characterization of the string, 
in terms of the number of 7-branes piercing the D3 brane wrapped on $B_Z$ \Witpt.  
For the manifolds $Z_n$ with $n=0,1,2$ this number is $16,8,0$
and one gets the current algebra of a  heterotic string, a non-critical string with $E_8$ current algebra, 
and a $N=2$ supersymmetric string that is equivalent to a D3-brane wrapped on 
the 2-sphere of an $A_1$ singularity in K3, respectively \Witpt.

The section $B_Z$ of $Z_0$ can not be contracted, in agreement with the non-zero tension of the 
heterotic string, but the strings with $n=1,2$ can get (classically) a zero tension at Vol$(B_Z)=0$
\Witc\MV\SWts.%
\foot{It is amusing to note that upon compactification to 4d, the non-zero tension of the heterotic string 
becomes equivalent to the statement that the pure $SU(2)$ $\cx N=2$ gauge theory in 4d associated
with shrinking the section $B_Z$ is always on the Coulomb branch \SW. Moreover the non-zero tension of 
this heterotic string is exponential in the volume of the second $\IP^1$ in $F_0$ and is in fact a
non-perturbative effect, see also the discussion in \PMff.}

The 6d F-theory compactification on $Z$ is related to the 4d type IIA compactification by a 
further compactification on $T^2$. An Euclidean wrapping of the 6d string on $T^2$ gives rise
to a world-sheet instanton contributing to the superpotential
\defWi. The lowest mode contributes the first term in 
\ismodi, but there is an infinite number of more
massive contributions as is familiar from the 
heterotic string. To illustrate this point, consider the 5d particles obtained by 
a once wound string on $S^1$ and carrying KK momentum along the circle. These states can be related 
to world-sheet instantons of the type IIA string \KMV. Let us see how the world lines of 
these particles wrapping the second $S^1$ contribute to the
superpotential \defWi. The answer can be computed from the 
Gromov-Witten invariants of the type IIA string and it is
\eqn\winsts{
\ffi^{Z_n}=\cases{b_0\, Q\, (1 - 240q_1- 141444q_1^2+...)=b_0\, Q\, \fc{E_4E_6}{\eta^{24}}(q_1),&$n=0$,\cr
b_1\, Q\, (1 + 252q_1 + 5130q_1^2\ \ \ +...   )=b_1\, Q\, \fc{E_4}{\eta^{12}}(q_1),&$n=1$,\cr
    b_2=0&$n=2$,}
}
valid up to $\cx O(Q^2,q_2^1)$.\foot{Here $q_i=\exp(2\pi i t_i)$ as previously and 
$t_1\sim U$ and $t_2\sim\SS$  are the two 7-brane moduli;
see app.~A.1 for more details.} The expression for $n=0$ is the supersymmetric index for 
K3, which plays a prominent role in the 1-loop computations of the heterotic string \HM.
The expression for $n=1$ is the partition function of the non-critical $E_8$ string \KMV.
Finally the zero for $n=2$ is a consequence of the zero beta function
coefficient and it does not say that this string is trivial. In fact the manifold $Z_0$ is a
deformation of $Z_2$ in the complex structure \MV, corresponding to giving a vev to the adjoint
hyper multiplet in the 4d gauge theory with beta function coefficient $b_0$ in \beta. 
So this zero of the superpotential
should be thought of as a zero of the determinant factor at a special value of 
the complex structure, as discussed in \Witsp. 

Thus the various dual pictures of the instanton generated superpotential
all nicely fit together:
the 1-loop effects on the world-volume of the M5-brane,
the D3-instantons in F-theory related to the world-sheet instantons of a  (non-critical) 6d string
and the world-sheet instantons of the dual type IIA string on \CY\ manifold $Z$. 
It would be interesting to study these relations in more detail.
To study the vacuum structure of the flux compactification, we will rely on the type II picture
in the following.

\subsec{Adding space-filling D3-branes}
The above compactifications are dual to F-theory compactifications without space-filling D3-branes.
We will now study some simple aspects of the case with D3-branes included. 

We will only address the two most basic questions and consider them in a minimal compactification: 
$i)$ what is the perturbative effective supergravity including the open string modulus, $\BB_\al$,
that measures the position of the D3-brane relative to the 7-branes. $ii)$ what is 
the instanton corrected superpotential for the D3-brane modulus. 

\def\Zbu{Z_1^{*}}
As discussed above, a space-filling D3-brane can be added by blowing up a point 
in the base of the elliptic fibration $Z_n$, which is the Hirzebruch surface $F_n$.
As an example we consider a blow up of $Z_1$ along the exceptional section of $F_1$.
The resulting manifold $\Zbu$ has four scalars $t_a$ that describe two independent 7-brane positions $C_a$, the
D3-brane position $\BB$ on the base of the fibration $\kf \to B_V$, and in addition the volume modulus $\TT$ of $\kb$. 
The geometry of the manifold $\Zbu$ is described in app.~A.2. 

\subsubsec{Perturbative couplings of D3- and 7-branes}
Let us first study the classical supergravity action, related to the polynomial part
of the prepotential $\cx F$ in \cypp. In a compactification on $T^4$, 
D7- and D3-branes are related by T-duality along the 4 world-volume directions of the
7-brane and this should be reflected by a symmetry in the effective action. This 
argument has been used also in the construction of effective supergravity actions of 
toroidal orientifolds \ADFT\AFT.\foot{See also the related works in \LST\IBW\IBWii\LJ.} 
However we do not expect such a symmetry in a moderately generic case for the obvious reason that
the gauge theories on the D7-brane and the D3-brane have different global supersymmetries
and low energy dynamics.\foot{In fact, T-duality of a system of 
7-branes on K3 is not fully understood because of the non-trivial 
dynamics of the world-volume theory.} The 1-loop 
corrections to the coupling are different and therefore we expect an asymmetry in the couplings
even at the perturbative level; it turns out that the 1-loop correction is
significant.

There is a certain ambiguity in how one parameterizes the relative positions of the 3 branes on $B_V$; 
the coordinates on the K\"ahler cone correspond to one relative 
7-brane position and the distances of the 3-brane to the 
two 7-brane but one may prefer different coordinates.\foot{This is similar to an ambiguity in the
heterotic string duals, see e.g. \LSTY.}  
The ambiguity will affect the perturbative couplings of the D3-branes to the 7-branes, 
as it corresponds to expanding around different vevs for the scalar fields.

\def\pari{(A.4)}
A simple dimensional reduction leads to the couplings on the brane world-volumes 
\eqn\drcps{
\fc{1}{g_{D7}^2}=\Im \TT,\qquad \fc{1}{g_{D3}^2}=\Im \SS,
}
which can be generalized to holomorphic functions as in \GGC. Let us compare this with the
result from the dual type II string on $Z$.
We choose a parametrization where $\BB=0$ corresponds to a D3-brane at a generic 
position away from the 7-branes;\foot{For another parametrization adapted to the D3-brane
sitting on top of a 7-brane see (A.5).} the relation to the K\"ahler moduli of $\Zbu$ is given in \pari.
This leads to the cubic prepotential
\eqn\preppp{
\Fc = \TT\ (\SS U) -\fc{1}{2}\SS \BB^2 +\fc{7}{24} \SS^3.
}
Although this result is apparently consistent with \drcps, there are surprising details.
Without the last term the prepotential \preppp\ agrees with those for the
supergravities based on certain homogeneous spaces, discussed in \ADFT\AFT. However there 
is in addition the $\SS^3$ coupling that corresponds to a 1-loop effect from the point of the 7-brane world volume.
It leads to a further mixing of the fields $\SS$ and $\BB$; this asymmetry is expected 
from the lower supersymmetry on the 7-brane. 

More importantly it follows from the parametrization \pari\ that the above result for the 
D3-brane coupling is only valid in a regime in the moduli where 
$$
\Im U> \Im \SS,
$$
which is not the standard regime of the perturbative type II string. 
In other words, the classical gauge coupling on the D3-brane is the {\it smaller} of the two 
7-brane moduli. 

In the usual derivation of \drcps, one assumes $\Im U$ and $\Im \SS$ to be large, but one does
not specify the ratio of the two. However it follows from the heterotic dual, that
the type IIB on the orientifold \ofi\ has a strong coupling singularity at
$\SS=U$. The perturbative couplings depend on the sign of $\Im U-\Im \SS$. For $\Im U \sim \Im \SS$
there are large perturbative as well as non-perturbative corrections that violate the
relation \drcps\ badly. Continuing to the regime $\Im U \ll \Im S$ the theory is 
again weakly coupled, but the bare coupling on the D3-brane is $\Im U$, not $\Im \SS$. 
This is relevant also for studies of flux superpotentials in orientifold models.

Another consequence of \preppp\ is that the modulus $\TT$ mixes with
the other moduli already at the perturbative level (this is even true in the
case without D3-branes), through the K\"ahler potential
\def\SSb{\bar{S}_{II}}
$$\eqalign{
K=&-\ln\big(\fc{i}{24}(\SS{-}\SSb) (7(\SS{-}\SSb)^2{-}12(\BB{-}\bb \BB)^2{+}24 (\TT{-}\bb \TT)(U{-}\bb U)){+}
\fc{105}{\pi^3} \zeta(3)\big)\cr
=&-\ln(i(\TT-\bb \TT)(U-\bb U)(\SS-\SSb))\cr
&+\fc{1}{(\TT-\bb \TT)(U-\bb U)}\big(\fc{1}{2}(\BB-\bb \BB)^2-\fc{7}{24}(\SS-\SSb)^2+
\fc{105i\zeta(3)}{\pi^3 (\SS-\SSb)}
\big)+...
}$$
where the second term in the second line is the 1-loop correction
to the K\"ahler potential, in the expansion of the 7-brane gauge coupling.
For comparable values of the moduli, the above K\"ahler potential deviates considerably
from the factorized form assumed in the study of many orientifold models.

\subsubsec{Non-perturbative superpotential for D3-brane moduli}
As can be seen from \preppp, and argued more generally from the geometry of the
dual \CY\ manifold, the D3-brane moduli $\BB_\al$ do not couple
to the K3 volume modulus $\TT$ in the classical prepotential $\cx F$.  
The absence of these couplings to $\TT$ means that the D3-brane moduli $\BB_\al$
do not appear in the classical 
period vector \newbas\ and do not enter the classical flux potential $W_F$ in \wf,
at least for the subset of fluxes in \Fflux.
The moduli $\BB_\al$ may therefore enter the superpotential only via the instanton
corrections, and behave in this sense very much like the K3 volume modulus, $\TT$.

We will now show that the instanton corrections indeed depend on the 
modulus $\BB$ in a generic enough way to fix all these moduli 
by the non-perturbative potential. To do so it is sufficient to
study the first terms of the potential computed in the dual type IIA string. 
In an expansion for large K3 volume and at weak coupling, $\Im \SS,\Im \TT \gg 1$, the first terms are
\eqn\winstdt{
\eqalign{
\ffi=
&q_4 \ \big( \fc{q_3(1+\qh_2)-2\qh_2}{(1-\qh_2)^2})\cr
&+q_4\qh_1\, (252q_3+\fc{420\qh_2(1+\qh_2-2q_3)}{(1-\qh_2)^2})\big)
+\cx O(q_4^2,\qh_1^2).
}}
Here $\qh_1=\langle q_1 \rangle$ and $\qh_2=\langle q_2q_3\rangle $ denote 
small numbers that correspond to the values of the 7-brane moduli fixed by the classical 
potential, $W_F$.  The above instanton correction 
shows a rich, generic  dependence on the K3 volume modulus $Q=\exp(2\pi i \TT)$ and 
the D3-brane modulus $q_B=\exp(2\pi i \BB)$ through the exponentials
$$
q_4=Qq_B^{-1/2}\, (\qh_1\qh_2)^{-7/8},\qquad q_3=q_B (\qh_2)^{1/2}.
$$ 
These flat directions of the classical potential will be fixed by the non-perturbative potential \winstdt.
Note that this will generically happen at a point in the moduli where the D3-brane is
fixed away from the 7-branes. It would be interesting to study the mass scales
on the 7-brane world-volume generated for the matter multiplets from the 37-strings in detail.

\subsubsec{More branes}
For the identification of the manifold $\cx M_V$ for the effective supergravity, 
it is instructive to consider a slightly more involved case with one more 7-brane.
For completeness we record here the general form of the cubic prepotential for 
the type II dual described in app.~A.3:
\eqn\fvmod{
\Fc=
\TT\, (\SS U-C^2)+(-\fc{1}{2} \BB^2 \SS-\SS^2 U-\fc{1}{2} \SS U \BB
+\fc{7}{24} \SS^3+\fc{4}{3} C^3-\fc{1}{8} \SS U^2)\ .
}
Here $(U,\SS,C)$ denote the 7-brane moduli and $B$ the D3-brane modulus.
The above expression shows a rich structure of perturbative corrections
in the various coupling constants. 

\subsec{Remarks}
The instanton correction \defWi\ has an interesting dependence on the complex structure
moduli of the two K3 factors, which viewed from the 5-brane world-volume arises as a 1-loop effect \Witsp\Ganor. 
It would be interesting to develop the microscopic computation of the D3 instantons from a 
world-volume perspective and the explicit results obtained here may be of help for progress in this question.
It may be worth noting that the instanton corrections to the K\"ahler potential, which 
are the original D3-instantons with at most 4 and 8  zero modes, are included in the type II result
as well.

One remark on line bundles.
The effective $\cx N=1$ superpotential is defined as the holomorphic section of a 
certain line bundle $\cx L$, defined {\it classically} by the holomorphic (4,0) form on 
the 4-fold $X_4$ for the compactification. In the present geometry, this line bundle splits into
two line bundles $\cx L_H$ and $\cx L_V$ of the holomorphic $(2,0)$ forms on $\kb$ and $\kf$, 
with curvatures determined by the
K\"ahler potentials on (a submanifold of) 
the scalar manifold $\cx M_H$ for the hyper multiplets and 
the scalar manifold $\cx M_V$ for the vector multiplets, respectively. This will play an important
role in sect.~6. Note that the coefficient $c_k$ of the $k$-th instanton term $Q^k$ is 
{\it not} the section of the line bundle of the superpotential, as is sometimes claimed in the literature;
however, it is the section of a different line bundle.
The section defined by the superpotential includes the moduli dependence on $\TT$ through the instanton exponential.
As will be discussed briefly in sect.~6 this line bundle $\cx L$ is affected by further quantum effects.

\newsec{Generalized type II compactifications}
In the previous part of the paper, we were using duality in one direction, namely to learn about the 
non-perturbative F-theory potential from the closed string dual. In this section we study the
consequences in the other direction, leading to a natural proposal for a generalization of the 
Gukov-Vafa-Witten superpotential to the case of generalized type II compactifications.\foot{Related
results have been obtained in \GLW\ from a study of generalized CY manifolds ($cf.$ \Hit).}

\subsec{Generalized GVW superpotential}
The popularity of flux compactifications is
largely due to the fact that they allow to understand non-supersymmetric ground states
as the vacua of a supersymmetric compactification with supersymmetry spontaneously broken by the flux.
We have used this fact before, when computing the instanton corrections to the superpotential,
essentially from a different amplitude in the parent theory with higher supersymmetry.

The search for the most general gauge and geometric backgrounds in string theory 
that can be described within an underlying  $\cx N=2$ effective supergravity is still 
ongoing \vafaln\GLMW\Ruben\LGM\KaP\FerH\GLW. 
The effective supergravity gives an expression for the scalar potential of the theory in terms
of gaugings of the isometries of the manifold of scalar fields \thebible.
However it turns out that for many purposes it is much simpler and more powerful to use
an $\cx N=1$ language. For the case of gauge 3-form fluxes in type IIB, one has the
well-known Gukov-Vafa-Witten superpotential \GVW
\eqn\gvw{
W=\int \Omega \wedge G_3,\qquad G_3=F_3-\SS\,  H_3,
}
where $F_3$ is the RR 3-form and $H_3$ is the NS 3-form.

Generalized type II string compactifications involving metric and dilaton degrees
of freedom turn out be quite complicated technically. The simplest case is the
mirror of the type IIB theory with both RR and NS fluxes as described by \gvw, 
and it leads to compactifications on generalized (half-flat) manifolds discussed in \vafaln\GLMW.
These are non-complex manifolds, whose geometry is less manageable than 
that of Calabi--Yau manifolds. It would be very helpful to have a simple superpotential 
that generalizes the GVW superpotential to these cases. 

We will now propose such a generalization. On a subset of fluxes specified below, 
the proposal can be explicitly derived
from a chain of string dualities; since the argument requires some geometrical background
we give first a 
physical argument and defer a more rigorous derivation and some details to the next section.

The GVW superpotential \gvw\ has an elegant physical interpretation as the tension of 
4d domain walls obtained by wrapping the RR and NS 5 branes of the type IIB on 
3-cycles in the manifold. This relation to domain walls suggests 
that the generalized superpotential should also be simple. In fact, the F-theory superpotential \wf,
\eqn\wfagain{
W_F=\sum_{I,\La}\ \Pi^I(q^i)\, \g_{I\La}\, \Pit^\La(t_a),
}
is also related to the (classical) tension of 4d BPS domain walls, obtained from a 5-brane
stretched between the 7-branes on the upper K3 $\kf$ and wrapping a 2-cycle in the lower
K3 $\kb$. The 4d BPS tension is proportional to the wrapped volume and this is measured 
by the bilinear combination of the periods of the $(2,0)$-forms in $W_F$.

\def\II{{\bf I}}\def\LL{{\bf \Lambda}}
We will now argue that the natural expression that describes the exact BPS tension of these
domain walls in terms of the  dual type II side is also a simple bilinear, of 'quantum periods':
\eqn\wgen{\eqalign{
W(\Zb)=&\sum_{\II=1,...,h^{odd}(Z) \atop \LL=1,...,h^{even}(Z)}
(\int_{\ga_\II}\, S_{II}\,  \Om)\ \g_{\II\LL}\  (\int_{\al_\LL}  e^{B+iJ})\cr
=&\sum_{\II=1,...,h^{3}(Z) \atop \LL=1,...,h^{3}(\Zh)} 
(\int_{Z} S_{II}\,  \Omega \wedge \ga^\II)\ \g_{\II\LL} \ (\int_{\Zh} \hx \Omega \wedge \gah^\LL).
}}
The idea of the above formula is that this superpotential describes the type II compactification on a generalized 
(non CY-)manifold $\Zb$, which is related to its CY parents $(Z,\Zh)$ by switching on quantized 
background fields induced by the domain walls. Here $\{\ga^\II\}$ ($\{\ga_\II\}$) is a basis for   
$H^3(Z,\IZ)$ ($H_3(Z,\IZ)$) and a similar notation is used for the mirror $\Zh$. Moreover $\{\al_\LL\}$ is a basis for 
$H_{2k}(Z,\IZ)$.

Before continuing, a few words of caution. In the above expression we use 
bold-face letters $\II,\LL$ to denote indices that run over the full cohomologies, 
whereas ordinary letters will be reserved for a set of ``electric'' $A$-cycles
under the symplectic pairing in $H^3(X,\IZ)$, running over $\fc{1}{2}h^3$ indices.
The derivation in the next section applies strictly only to a subset of the electric gaugings,
corresponding to the ordinary indices $I,\La$ in \wfagain. We will give some plausible arguments
that \wgen\ is the natural generalization in the simultaneous presence of 
electric and magnetic charges, but the generalization is non-trivial. Moreover
one expects that there will be some consistency conditions on the matrix $\g_{\II\LL}$,
generalizing the tadpole condition $\int F\wedge H=0$ of \TV (see also \KaP). 

With this preliminaries and up to a small modification that will be added momentarily, 
eq.\wgen\ is our proposal for the generalized GVW superpotential describing 
general backgrounds induced by a large class of new domain walls in the type II theory.
The relation between the deformed background $\Zb$ and its parents $(Z,\Zh)$ 
is similar to the case of  half-flat manifolds, which are not \CY, yet the superpotential 
is computed by the quantities in a parent \CY\ \vafaln. 
In other words, the superpotential is the section of a certain line bundle,
which is made from contracting the two symplectic sections defined by the holomorphic (3,0)
forms on the parents $(Z,\Zh)$. The background determines the contraction, but it does
not change the geometry of the underlying bundles.\foot{It would be interesting
to relate this contraction to the framework of \Hit.}

Although the formula is obviously mirror symmetric, we consider it in the context
of type IIA theory on $Z$ for concreteness. 
As already indicated by the second expression, the 
integrals above should be thought of in terms of quantum cohomology. The 'quantum 
integrals' of $e^{B+iJ}$ contain the world-sheet instantons of the type IIA theory (which
played an essential role in the foregoing sections), and these are summed up by the 
period integral on the mirror $\Zh$. Similarly the period integral on the hyper multiplet
side should be generalized to include all quantum effects.

Before discussing these quantum effects, let us connect the above superpotential
to the quantities in the effective $\cx N=2$ supergravity. The scalar manifold 
$\cx M_H$ for the hyper multiplets of  type IIA string compactified on $Z$ is a quaternionic space of 
a special type, called dual quaternionic in \refcmap. The number of hyper multiplets is
$n_H=1+h^{12}$, where $h^{12}$ is the number of holomorphic (2,1) forms on $Z$. The 
$4n_H$ scalars in the hyper multiplets of type IIA theory compactified on $Z$ arise as 
$$\eqalign{
h^{21}\ :&\ z_a,\ \xi^a\sim C_{ij\bb k},\ \xit_a\sim C_{\bb i\bb j k},\cr
1\ :&\ \phi,\ a,\ \xi^0\sim C_{ijk},\ \xit_0\sim C_{\bb i\bb j\bb k},
}$$
where $z_a$ are the complex structure moduli of $Z$, $C$ the RR 3-form,
$\phi$ is the dilaton and $a$ the 4d scalar dual to $B_{\mu\nu}$. The space $\cx M_H$
has (at least) $2n+3$ Abelian isometries which can be gauged under
the $U(1)$ gauge symmetries. These correspond to shifts of the $2n+2$ fields $\xi^A,\xit_A$, $A=0,...,h^{12}$
and the shifts of the NS axion $a$ \refcmap.

The most general (hyper multiplet) gaugings in supergravity have been considered so far in \FerH, 
where it was shown that one can gauge the $2n+3$ isometries under certain consistency conditions 
on the Killing vectors that follow from the underlying Heisenberg algebra defined by the isometries.%
\foot{These conditions should be related to the generalized tadpole condition mentioned before.} 
The formula $\wgen$ is the generalization of the GVW superpotential for these more general 
supergravity gaugings, going beyond the 3-form flux $F_3-\SS H_3$ in type IIB language.
The charges of the scalars in the hyper multiplets under the $h^3(\Zh)=2n_V+2$ electric and magnetic
$U(1)$ symmetries are given by the integral matrix $\g_{\II\LL}$; for the electric gaugings
in F-theory this fact is derived in \fres. However note that the matrix $\g_{\II\LL}$ has only 
$h^3(Z)\times h^3(\Zh)=(2n_H+2)\times (2n_V+2)$ entries, and thus misses $2n_V+2$ independent
gaugings of a single hyper multiplet scalar under the $U(1)$ gauge symmetries. 

The missing independent gaugings are related to the isometry of  
the NS axion $a$. In \Mich\ ($cf.$ \PS) the supergravity associated with 
the GVW superpotential has been studied, and it was shown that the gauging of $a$ arises 
from the RR 3-form flux $F_3$. That is,
although \wgen\ is on the one hand a substantial generalization of the GVW superpotential, 
it misses the piece $\int_Z F_3\wedge \Omega$ due to the RR flux.

The missing of this term in \wgen\ is not essential for a background that preserves the 10d $SL(2,\IZ)$ invariance
of the type IIB string,  as one can always rotate to a basis where there is only a NS flux.  
On the other hand, backgrounds that break the 10d $SL(2,\IZ)$  can be described by 
the replacement 
\eqn\piex{
\Pi^\II\, \g_{\II\LL} \longrightarrow \Pi^{\II}\,  \g_{\II\LL} + N_\LL,
}
in \wgen, where 
$$
\Pi^\II=\int_{\ga_\II} S_{II}\,  \Omega
$$
denotes the period integrals in the hyper multiplet sector of the type IIA theory.
This leads to the extra term
\eqn\wex{
\delta W(\Zb)= \sum_\LL N_\LL \ (\int_{\Zh} \hx \Omega \wedge \gah^\LL),}
which describes the superpotential for 
a RR flux $F_3=N_\LL \gah^\LL$ in the type IIB theory.

The presence of the shifted terms \piex\ will be argued more rigorously in the next section;
a physical interpretation is as follows. 
As alluded to above, the extra term is really related to backgrounds that break 
the 10d $SL(2,\IZ)$ invariance. To motivate the replacement \piex,
as well as get further insight into the formula \wgen, recall that
the quantum definition of the period integrals $\Pi^\II$ is as
objects in the topological string theory on $Z$. It was argued in \NOV,
that the topological string has an \S-duality, which is inherited from the 
10d $SL(2,\IZ)$ duality from the type II string. It was further argued that there
is a coupling
\eqn\sdc{\eqalign{
\int \fc{(\Omega_B +ig_B^2 C_{NS})\wedge d(k_B+ig_B B_{RR})}{g_B^2} 
\matrix{S\cr\leftrightarrow\cr\phantom{S}}\int \fc{(\Omega_A +ig_A C_{RR})\wedge d(k_A+i B_{NS})}{g_A^2} 
}}
where the subscript $A$ ($B$) refers to the A-model (\S-dual B-model) on
a CY, $g_A=g_B^{-1}$ are the topological coupling constants, and $C$ and $B$ are 
3-form and 2-forms, respectively.

Note that the above coupling reproduces both, the NS and RR term in \gvw, 
from the terms proportional to $dB_{NS}$ and $dB_{RR}$ but to describe 
these terms with general coefficients that break the $SL(2,\IZ)$ one has to add in some sense
both terms at the same time. Note also that \sdc\ defines the superpotential
for the case of type IIA compactified on the special half-flat manifolds  considered in \GLMW.

The antisymmetric tensor fields $B$ and $C$ are defined up to discrete, integral shifts
by elements of the cohomology.\foot{This is oversimplified, see e.g. 
\Morr. Moreover, whereas the closed string sector is
almost invariant under these shifts, the open string sector is not.}
The breaking of the $SL(2,\IZ)$ is a quantized effect
and it is natural to think of it as shifting the values of, say the NS field relative
to the RR field by an integer, in a way that breaks \S-duality. In fact shifting, say in IIA,
$$
C_{RR} = c_\II\ga^\II+ \cx S(C_{NS}),\qquad c_\II\in \IZ,
$$
where $\cx S$ denotes the \S-duality operation, the coupling
\sdc\ produces precisely shifts  of the form \piex.

Note that, similar as the original F-theory expression \wfagain,  the superpotential is a section of 
two separate line bundles over the manifolds $\cx M_V$ and $\cx M_H$. This factorization is 
preserved by the instanton corrections discussed in the previous sections.

\subsubsec{T-dualities}
Note that integral basis transformations of $H^3(Z,\IZ)$ and $H^3(\Zh,\IZ)$ 
generate a large group of ``$T$-dualities'',
\eqn\tdual{
\cx G_T=\Sp(h^3(Z),\IZ)\times \Sp(h^3(\Zh),\IZ).
}
These dualities can lead to symmetries in the moduli of a single theory, if the
transformation is induced by a monodromy in the original moduli space and
at the same time the flux matrix respects these symmetries. Otherwise the
duality relates the effective theories of different compactification backgrounds. 
This is similar as in the more familiar case of the moduli space
of $\cx N=2$ parent theories. A continuous, real version 
of \tdual\ is realized in the classical supergravity \FerH.

The above transformations do not mix with the extra component \wex\ related to the
RR flux. It is natural to ask, how the $SL(2,\IZ)$ transformations that
act on the GVW superpotential \gvw, combine with the T-dualities in \tdual. 
In the next section we will give evidence, that the combined transformations
generate a duality group
\eqn\sdual{
\cx G_S=\Sp(h^3(Z)+1,\IZ)\times \Sp(h^3(\Zh),\IZ).
}

\subsubsec{Statistics of vacua}
Taking into account that the final flux matrix is an integral
$(h^{3}(Z)+1)\, \times\,  h^{3}(\hx Z)$ matrix, 
the counting of flux vacua in generalized type IIA compactifications with tadpole $L$ 
should be modified to  
$$
N_{vacua} \sim L^{\fc{1}{2}\, (h^{3}(Z)+1)\, \times\,  h^{3}(\hx Z)}.
$$
replacing the estimate $N_{vacua} \sim L^{2(n_V+1)}$ based on the GVW superpotential \AD.

\subsec{Derivation of eqs.\wgen\ and \wex}
In this section we will derive the formula \wgen\ and the shifts \piex\ in the limit of
small type IIA coupling constant and for a restricted set of fluxes. 
The argument is based on a detailed study of the string duality
considered in this paper
and involves essentially three ingredients, the c-map of \refcmap,
the `stable degeneration' (s.d.) limit of \MV\SDii, and a certain result on mirror
symmetry obtained in \BM. The general idea is to first use that the lower K3, $X_H$, is the same for the 
F-theory and the heterotic dual on $T^2\times K3$. In the second step
we use a very explicit map from the complex structure moduli of 
heterotic on $T^2\times K3$ to type IIA on $Z$, which goes via the 6d F-theory
in the s.d. limit.

\def\kbe{X^{e.f.}_H}
Let us first sketch the argument for a codimension two subslice in the F-theory fluxes. 
Consider a non-generic heterotic K3, $\kbe$, of the same type as the upper F-theory K3.
That is $\kb$ is an elliptically fibered, polarized K3 with a Picard lattice of dimension
2 spanned by the class $E$ of the elliptic fiber and the class $B$ of the base of the fibration
$\kbe \to B$. The complex structure moduli of $\kbe$ preserving this structure parametrize
the coset \vafaln
\eqn\cscoset{
SO(\Gamma^{2,18})\backslash SO(2,18)/SO(2)\times SO(18).
}
The lattice $\Gamma^{2,18}$ corresponds to 20 out of 22 2-forms $\eta^I$ on $\kb$
that are available for the F-theory flux \Fflux\ entering the superpotential. Two fluxes
have been lost because of the holomorphic elliptic fibration (fluxes on $E$ and $B$
may contribute to the $D$-terms in \defvd).

There are two K\"ahler moduli of the elliptic fibration $\kbe$. The first is the
size $t_E$ of the elliptic fiber $E$. The limit $t_E\to \infty$ in the heterotic
string maps to the stable degeneration $\Zsd$ of the elliptic and K3 fibration 
$Z$ \MV\SDii. In this limit, the bundle data of the heterotic string on $\kbe$, encoded in the
complex structure of $Z$, are lost and the complex structure moduli of $\Zsd$
map 1-1 to the complex structure moduli of $\kbe$.
The size $t_B$ of the base $B$ of the fibration $\kbe \to B$ maps to the type IIA 
dilaton \AHM. 

What we want to show first 
is that in the limit of large $t_E,t_B$ one recovers the classical definition of
the CY periods from the K3 periods,
$$
\int_Z \Omega\wedge \ga^I\simeq\int_{\kb} \om \wedge \eta^I,\ \
$$
where $\simeq$ means up to perturbative symplectic rotations of the CY period vector.
Combining this with the identifications of the periods on the side of the vector multiplets, \fres,
this will establish \wgen\ for the subset of fluxes discussed above.

Some details on how the periods of K3 map to the periods of $\Zsd$ have been described in \PMade.
However for our purposes it is more convenient to use a result on mirror symmetry of ref.\BM.
It was shown there that if $Z_{n+1}$ is a CY $n+1$-fold for an F-theory compactification dual to 
the heterotic string on an $n$-fold $X_n$, then the mirror $\hx Z_{n+1}$ of $Z_{n+1}$ is a fibration 
$\hx X_n \to \hx Z_{n+1} \to \IP^1$ with $\hx X_n$ the mirror of the heterotic compactification
manifold $X_n$. Moreover the stable degeneration limit of $Z_{n+1}$ maps to the large base limit of 
$\hx Z_{n+1}$. The virtue of this 'theorem' is that we can write at once the period vector
of the manifold $Z$ in the s.d. limit:
\def\VV{V}
\eqn\perfac{
\Pi^\II(\Zsd)=\pmatrix{\ \ \Pi^I\cr \VV \Pi^I} \ .
}
The factorized structure of the period vector is characteristic of any fibration $X_n\to Z_{n+1}$ 
and can be seen as follows. In the mirror $\hx Z_{n+1}$, the period vector \perfac\ corresponds to
the complexified volumes of the even-dimensional cycles. In the limit of a large base 
with volume $\VV$, these are of the form $C_{2k(+2)}=C_{2k}(\hx X_n)\times (1,\IP^1)$ 
and their volumes are given by the period vector \perfac.\foot{More precisely,
one loses the cohomology associated with singular fibers in this approximation,
which includes the heterotic bundle data mentioned before.
See \PMff\ for more details on this argument including the intersection form on
the cycles. Note also that $\Pi^\II$ defines a 
symplectic section of the special form \newbas\ discussed in sect.~4. This is
a consequence of the fact, that in the M-theory compactification to 3d on a further
circle, there is a symmetry that exchanges the two K3 factors. This exchanges the
also the manifolds $\cx M_H$ and the hyper multiplet manifold $\cx M_V'$ associated
with $\cx M_V$ by the c-map.}

In our case, $n=2$ and the mirror $\hx Z$ of $Z$ is therefore a K3 fibration\foot{The K3
fibration of the mirror $\Zh$ should not be confused with the K3 fibration of $Z$ which
was required for having a perturbative F-theory dual in sect.~4.}, with 
a K3 fiber  $\hx X$ that is related to $\kbe$ by mirror symmetry of polarized K3 surfaces \Dol.
It follows that the upper half of \perfac\ is precisely the period vector $\Pi^I(\kbe)$ of $\kbe$.
Plugging the special form of $\Pi^\II$ and the F-theory fluxes  
into the general formula \wgen, we recover the F-theory
superpotential $W_F$ \wfagain, 
restricted to the fluxes that are available in the s.d. limit.

To recover the missing fluxes and to motivate the most general formula of sect~6.1,
let us see how the period $\Pi^\II$ of $\Zsd$ sits inside the manifold $\cx M_H$ for
generic values of $\VV$ and the type IIA dilaton. By the c-map \refcmap, the 
$4(n+1)$ dimensional quaternionic manifold $\cx M_H$ has a $2n$ dimensional 
K\"ahler submanifold $\cx K_H$ of special type. For the hyper multiplets
associated with a general K3 surface $\kb$ this relation is
\eqn\hkrel{
s^{-1}:\ \fc{SO(4,20)}{SO(4)\times SO(20)}\longrightarrow\big(\fc{SU(1,1)}{U(1)}\big)_V \times \big(\fc{SO(2,18)}{SO(2)\times SO(18)}\big)_{CS}\ ,
}
where $s$ is the basic map defined in \refcmap, arising from the compactification of the
type II string on a CY times an extra circle. The moduli on the l.h.s. of \hkrel\
are the hyper multiplet
scalars for the lower K3 $\kb$. As indicated, the first factor on the r.h.s of \hkrel\ 
can be identified with 
the complex modulus containing $\VV$ and the second factor with the complex structure of the 
specialized K3 manifold $\kbe$. This is in agreement with the factorized 
form of the period vector \perfac.

Note that a subset of the symplectic transformations acting on $H^3(Z,\IZ)$ acts as a $SL(2,\IZ)$
on the first factor on the r.h.s. of \hkrel; in particular there is one transformation exchanging
the upper and lower halfs of the period vector \perfac. Thus from the point of the CY geometry,
it is very natural to extend the set of gaugings to the full period vector $\Pi^\II$, as has been
implemented in \wgen.

To recover the extra term in \wex\ associated with the RR-fluxes in the type IIB dual, 
we have to consider a more general F-theory flux on a generic, non elliptically
fibered K3, $\kb$. This means necessarily to abandon one of the two K\"ahler moduli
$t_E$ and $t_B$, related to the s.d. limit and the type IIA dilaton, respectively.
To recover the classical limit, in which the 'quantum periods' $\Pi^\II$ are
related to the geometric period integrals we have to keep the type IIA dilaton large
and the volume of $\kb$ finite. 

We describe now a deformation to a generic K3 starting from the elliptically fibered case. 
Consider the K3 $\kbe$ with K\"ahler class 
$$
J=t_E E+t_B B,\qquad V=\fc{1}{2}\int_{\kbe}J\wedge J = t_E\, (t_E+t_B),
$$
at fixed K\"ahler moduli $(t_E,t_B)$. Here $(E,B)$ are a (dual) basis for $Pic(\kbe)$ with intersections
$$
E.E=2,\qquad E.B=1,\qquad B.B=0.
$$
Let $C$ denote the unique class orthogonal to $J$, normalized to $C.C=-2$,
$$
C=\fc{1}{\sqrt{1+\la}}(S-\la B),\qquad \la=\fc{t_B}{t_E},
$$
where $S=E-2B$ is the section the elliptic fibration. One can deform the complex structure 
to grow a component in the direction of $C$
\eqn\newdefo{
\mu = \int_C\om,
}
with $\om$ the $(2,0)$ form. Consider a series of the above deformations with decreasing $t_E$ but fixed volume $V$.
By the above identification of $t_B$ with the 4d dilaton, this is a weak coupling limit, $t_B\sim t_E^{-1}$, 
of the type IIA string.
We are interested in the scaling of the deformation \newdefo\ with the dilaton $t_B$.
{}From the definition of $C$ one finds that the deformation $\mu$ scales as 
$$
\mu \sim t_B.
$$
Adding the new period to the period vector $\Pi^I$ of $\kbe$ one obtains 
$$
\us \Pi^I=\pmatrix{g_{10}^0\ \Pi_{e.f}^I\cr g^1_{10}\ \mu'\ \ } + \cx O(g_{10}^2).
$$
Thus the new period vector $\us \Pi^I$ for 
the generic K3 $\kb$ has one entry of $\cx O(g_{10})$ that reproduces
the extra term \wex, and the potential from the RR-flux in the type II theory
is just one of the terms in the F-theory potential \wfagain. 
This is not surprising, since the type IIA dilaton maps to a hyper multiplet of the
F-theory K3. Note that the symplectic transformations on $\us \Pi^I$ mix the new period
with the original ones, leading to the generalized duality group \sdual.

Since the above limit requires small $t_E$, there are world-sheet instanton corrections 
to the hyper multiplet metric in the heterotic string. These are already included in the 
exact type IIA period integrals \perfac\ in the form of  
exponential corrections $\sim \exp(2\pi i V)$. See \pere\ for an interpretation of these instantons 
in the heterotic string. In fact these corrections are dual to the D3 instanton corrections in F-theory on a 
divisor of type $C)$ of sect.~2. It would be interesting to study this relation in more 
detail.

\subsec{Some remarks}
A cautionary remark is overdue at this point: it is not at all obvious, that the scalar potential
of the most general gauged $\cx N=2$ supergravity can be written in terms of a superpotential $W$. 
The Killing prepotentials $P^x_\La$, $x=1,2,3$ of the $\cx N=2$ effective supergravity are a triplet of the $SU(2)$ 
R-symmetry. To define the superpotential \spsugra\ one has to choose a preferred direction 
in the $SU(2)$ corresponding to the $D$-term. A careful discussion of how to define the $F$-terms
\wf\  and $D$-terms \defvd\ in the F-theory compactification can be found in \FF.
Note that the same flux gives rise to both terms. In the
general $\cx N=2$ scalar potential there are terms that mix the candidate $F$-terms
and the candidate $D$-terms and these mixed terms might be incompatible with the
$\cx N=1$ language. The ansatz \wgen, or even \gvw\ for a generic point in the hyper moduli,
does only apply to the case which can be consistently described in $\cx N=1$ terms. 

Secondly, we have not discussed the hyper multiplet moduli in the F-theory
that are present for general gauge bundles on $\kb$. From the point of the 
type II compactification on the CY, there is no clear distinction between these 
moduli and the geometric moduli associated with K3 and there is no reason
why one should restrict to the latter in the summation over the moduli of
$Z$ in \wgen. This is reasonable, but needs to be understood further. In particular
it would be interesting to understand the gauging of bundle moduli in the 
language of F-theory or the heterotic string.

Finally let us mention that the F-theory potential $W_F$ in \wfagain, has a large number
of classical supersymmetric ground states, both of $\cx N=2$ and $\cx N=1$ supersymmetry
\TriTri\FF. As discussed in the previous section, the F-theory fluxes map to 
quite general compactifications on the generalized CY manifolds of \Hit, 
where this property may be much less obvious. It would also be interesting
to study the vacuum structure of \wgen\ for the most general case. Note that
the most general superpotential \wgen\ depends on all CY moduli multiplets and should
generically lead to a complete fixing of all moduli even before taking into account 
instanton effects.

\newsec{Some remarks on the vacuum structure}
In this section we look into some properties of the vacuum structure 
of the instanton corrected scalar potential. We will not attempt a detailed
study, which will be considered elsewhere. 
We will restrict to point out a few interesting aspects that are worth of further studying. 

There are essentially two different cases, depending on whether the 
value of the superpotential at the classical vacuum, $W_0$, is 
zero or non-zero before taking into account instantons. 
In the no-scale approximation these correspond to supersymmetric and
non-supersymmetric vacua, respectively. We discuss these two cases separately.

\subsec{Vacua with $W_0\neq 0$}

The scalar potential for the F-theory compactification is given by the 
instanton corrected $F$-term \fullsp\ and the $D$-term \defvd. In \KKLT,
the minimization of the potential was performed in two steps, where one 
first fixes the vevs for the moduli appearing in the classical flux potential
$W_F$ and then uses the result as a potential for the moduli that
appear only in the instanton corrected part. From the results in \DK\ we expect, 
that this two step procedure is justified in a generic enough situation. However it is
important to check the absence of unstable directions, which is a necessary condition for 
the existence of an uplift to dS \CFNOP.

In the present compactification, the effectiveness of the classical potential $W_F$ 
in fixing most of the moduli has been discussed in \FF; see also  \AM\ for a similar
study for compactifications with D9/D5 branes. On the other hand the K3 volume $\TT$ 
and the D3-brane positions $\BB_\al$ do not appear in the classical flux potential 
and are to be  fixed by the non-perturbative effects. 

On general grounds, one expects that the F-theory compactifications considered
in this paper behave like the KKLT model for very large volumes. The advantage
of the present F-theory 
compactifications is that one can also compute the potential in the regime
of smaller volumes by string duality, thus probing a more generic class of string vacua. 
However, as we will discuss now,  also at large volume
there is a richer structure of vacua than in the KKLT model.

The reason is that the K\"ahler potential of the F-theory compactification
depends differently on the modulus $\rho$ that governs the instanton corrections.
Instead of being general we take as a simple example the compactification on $Z_0$ dual to F-theory without
D3-branes; the conclusions in more complicated models will be similar. The K\"ahler potential
is 
is $K=-\ln iY$, with 
\eqn\defYx{
Y=(t_1{-}\tb_1)\big(\fc{4}{3}(t_1{-}\tb_1)^2+(t_1{-}\tb_1)(t_2{-}\tb_2+\TT{-}\TTb)+
(t_2{-}\tb_2)(\TT{-}\TTb)\big)\ ,
}
where we have neglected $\al'$ and instanton corrections for simplicity.
As before, $\TT$ denotes the modulus which measures the K3 volume and 
determines the strength of the instanton corrections; on the other hand 
the moduli $t_1,t_2$ are generic D7-brane moduli that appear also in the 
classical flux potential $W_F$.

In the two step procedure, one looks first for solutions to the equations
$$
D_{t_a}W_F=(\p_{t_a}+K_{t_a})W_F=0,\qquad K_{t_a}=\p_{t_a}K,\qquad a=1,2.
$$
The problem with this is that the first derivatives $K_{t_a}$ depend on the K3 volume modulus $\TT$,
also for large values of $t_a,\TT$. Thus the value $W_0$ at a solution of the above equations
will be a function of $\TT$ and can not naively apply the two step procedure adopted in \KKLT.

Restricting the sizes of the moduli $(t_a,\TT)$ further, one can distinguish
two 
large volume regimes where the 
first derivatives $K_{t_a}$ are almost independent of $\TT$
\eqn\twor{
I: \Im \TT \gg \Im(t_a), \qquad II:\ \Im \TT \ll \Im(t_a)}
In these regimes, $W_0$ is approximately independent of $\TT$ and one can
attempt a two step procedure as in \KKLT. 

Despite of the fact that the K\"ahler potential is still different from that in the KKLT model in these regions,
the regime $I$ is expected to behave very similar. The other
large volume regime is different and leads to a large enhancement of the instanton
corrections. Let us recall why a very small $W_0$ was required in the KKLT ansatz and see
how this conclusion is relaxed in the regime $II$ of the F-theory duals. With a 
K\"ahler potential $K=-3\ln(\TT-\bb \TT)$ and the ansatz $\ffi=Ae^{ia\TT}$,
the minimization of the $\TT$ modulus 
\eqn\eDT{
D_\TT \big(W_0+\ffi(Q;\qh_a)\big)=0,\qquad D_\TT=K_\TT+\p_\TT,
}
reduces to 
\eqn\kkltmin{
W_0=-Ae^{-a\TT_2}\, (1+ia K_\TT^{-1})\ .
}
Here $\TT_1$ has been set to 0 for simplicity.
The function on the r.h.s. has a very small maximum for reasonable values of the
parameters $A$ and $a$, since the growth of the term $K^{-1}_\TT=2\TT_2/3i$ for large 
$\TT_2$ is cut off by the exponential suppression from the instanton prefactor.
A solution to this equation at sufficiently large positive $\TT_2$ thus 
requires a very small $W_0$ \KKLT.

With the K\"ahler potential \defYx\ one finds in the two regions 
\eqn\enhance{
K^{-1}_\TT \sim \cases{\Im \TT&$\Im \TT \gg \Im(t_a)$,\cr
\Im(t_a)&$\Im \TT\ll \Im(t_a)$.\cr}
}
We do not distinguish the values of the moduli $t_a$ for simplicity, with the understanding
that the precise behavior of $K^{-1}_\TT$ will depend on the details. The important point is
that for the non-minimal K\"ahler potential, $K_\TT^{-1}$ is considerably larger in the regime $\Im\TT \ll \Im t_a$
than in the KKLT model and this leads to a very effective enhancement of the prefactor of the instanton exponential. We have studied some examples and find that the enhancement factor can be
$\cx O(10-100)$. Fig.~1 shows the value of $W_0$ that solves $D_\rho W=0$ for the model
based on the manifold $Z_0$.

\vskip 0.5cm
{\baselineskip=12pt \sl
\goodbreak\midinsert
\centerline{\epsfxsize 2  truein\epsfbox{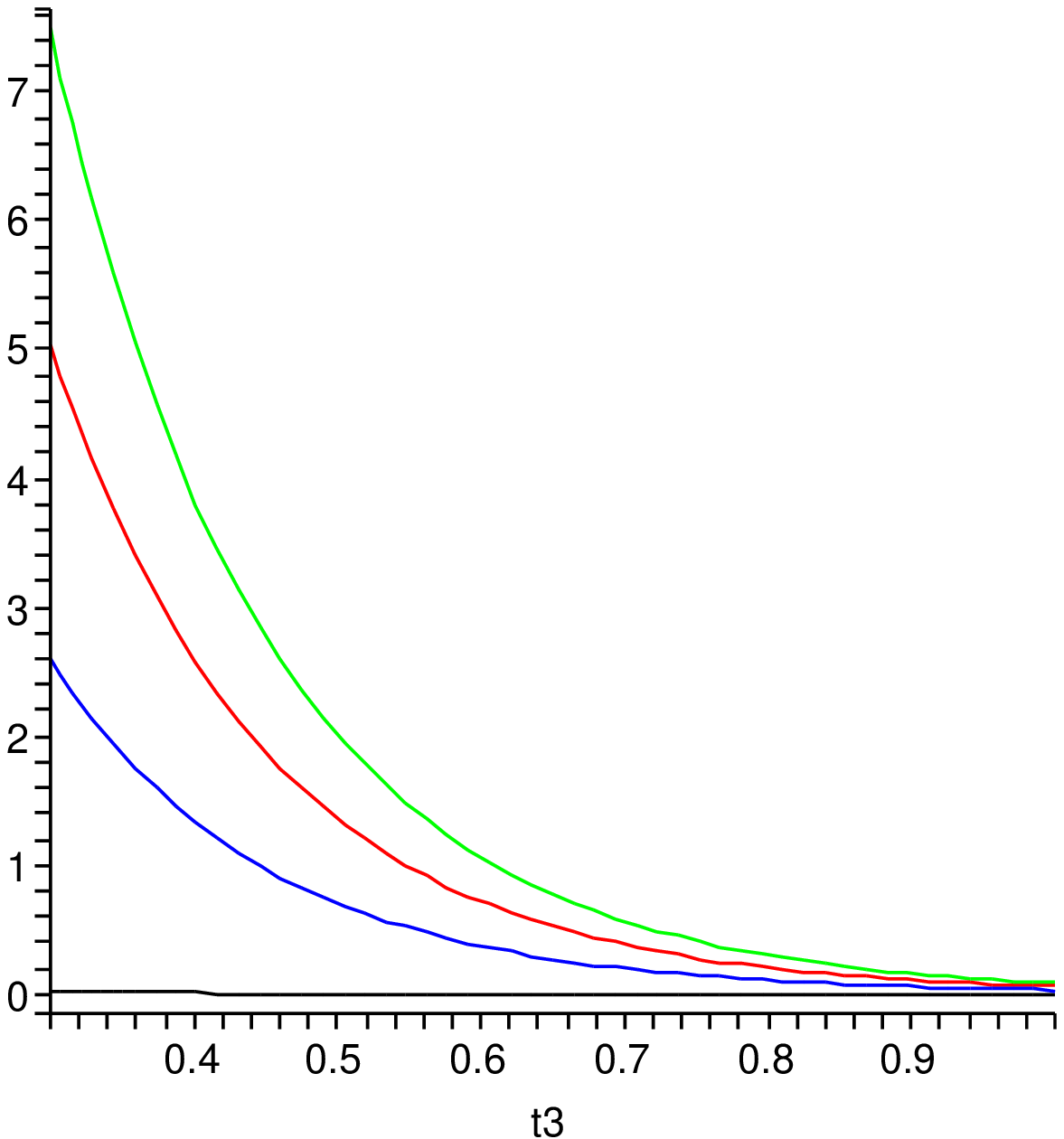}}
\leftskip 1pc\rightskip 1pc \vskip0.3cm
\noindent{\ninepoint  \baselineskip=8pt 
{{\bf Fig.~1:} Enhancement of the value of the classical superpotential $|W_0|$, 
as a function of $t_3$ and for a supersymmetric vacuum for $t_a=i(1+20n)$, $n=0..3$ 
from bottom to top.}}\endinsert}
\noi

Note that the enhancement does not depend on the details, and works if one or more of the $\Im(t_a)$
are sufficiently large. Although the value of $W_0$ is weakly correlated with the values of the $t_a$,
there is no reason to expect that the relevant ratio $K_\TT \cdot W_0$ is fixed
when increasing $\Im(t_a)$ in certain directions.
So the enhancement will take place for the  subset of fluxes\foot{One
expects that the selection is less serious for models with many moduli $t_a$, thus allowing
larger values of $W_0$. This is opposite to the case in \DDF, where many
K\"ahler moduli require even smaller values of $W_0$.}
 which lead to sufficiently large $\Im(t_a)$.

Note that it was also important here that there are two different types of K\"ahler moduli,
the $t_a$ fixed by the classical potential and $\TT$ fixed by the non-perturbative
effects. Otherwise the minimization equations for the $t_a$ would lead to 
equations similar to \kkltmin, but with the instanton prefactor being an exponential in $t_a$,
which would bring back the original condition $W_0\ll \cx O(1)$.

{\baselineskip=12pt \sl
\goodbreak\midinsert
\epsfxsize 2.6truein\epsfbox{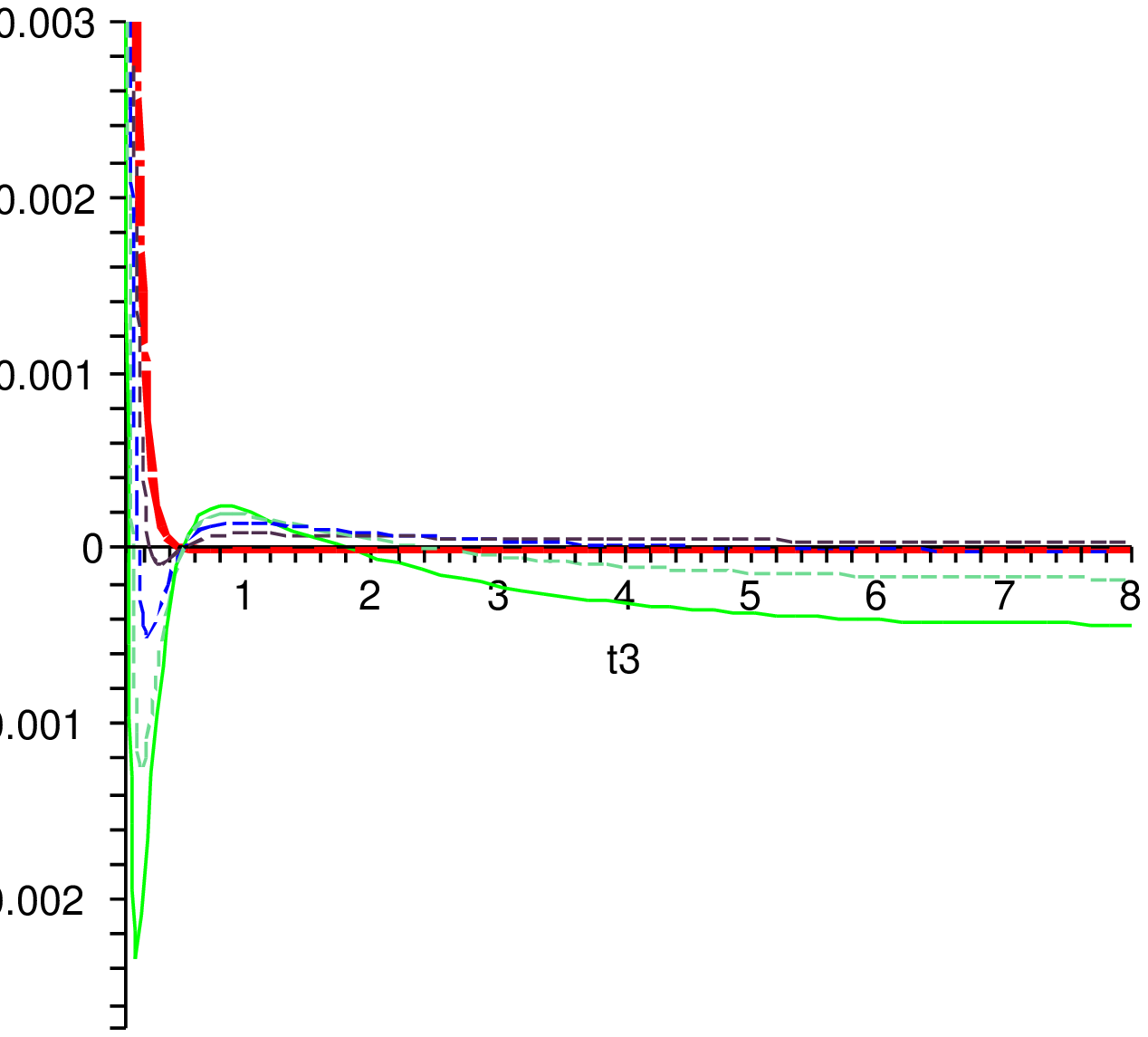}\hskip20pt
\epsfxsize 2.6truein\epsfbox{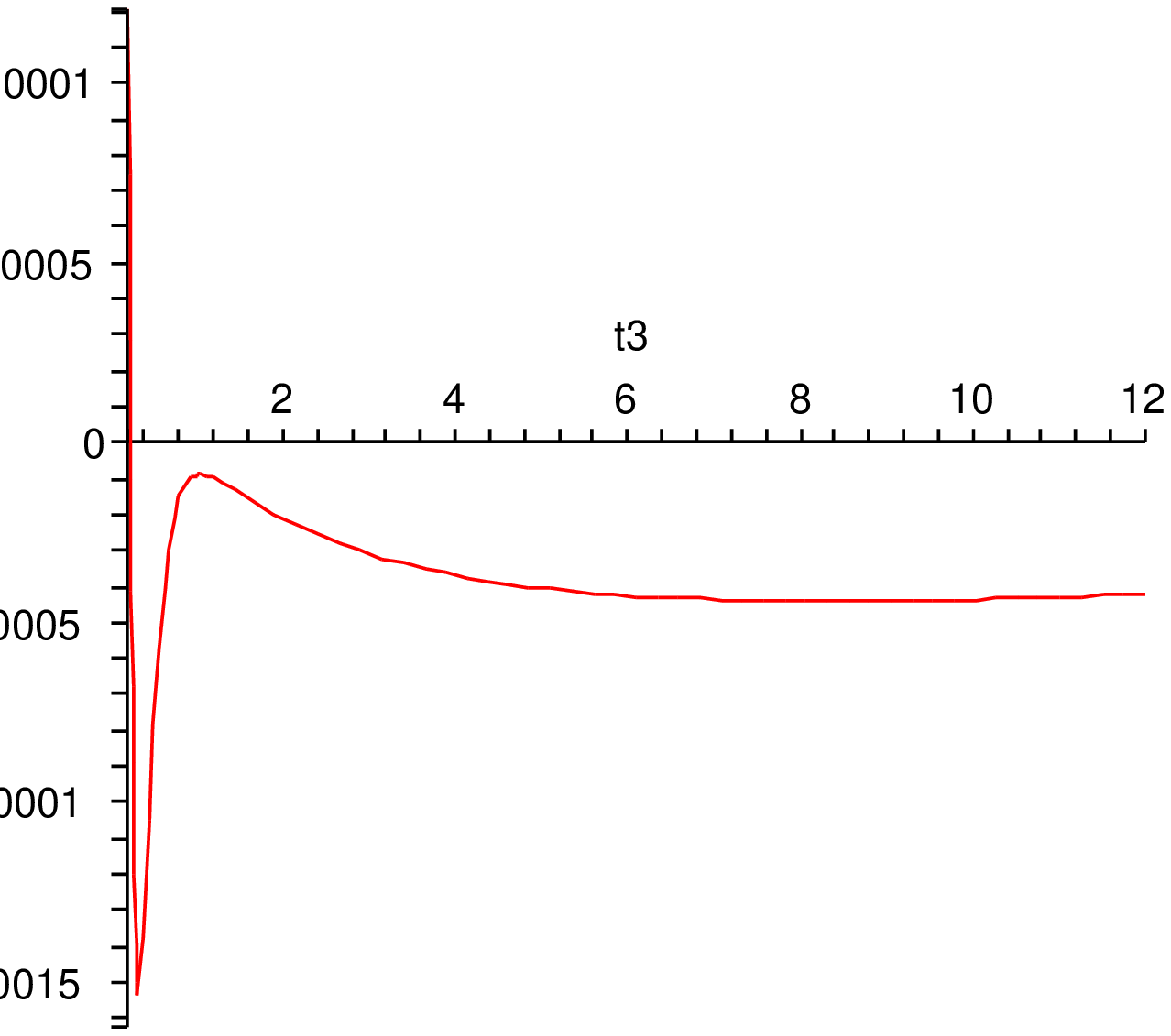}
\leftskip 1pc\rightskip 1pc \vskip0.3cm
\noindent{\ninepoint  \baselineskip=8pt 
{{\bf Fig.~2:} {\it Left:} Evolution of the $F$-term contribution to the 
scalar potential $V$ for increasing values of $|W_0|$. 
For larger $|W_0|$ the vacuum moves towards very small volumes whereas a new 
non-supersymmetric vacuum at large volume  appears above some $W_0^{crit}$. {\it Right:} Potential
with one supersymmetric and one non-supersymmetric minimum for $W_0>W_0^{crit}$.
}}\endinsert}

We have performed a preliminary study of the vacua, using the full instanton corrected K\"ahler potential
and superpotential, where the above enhancement can be explicitly studied. We find an interesting structure 
of the scalar potential with
either one or two minima, with and without supersymmetry. In particular there are non-supersymmetric
vacua similar to the ones discussed in \BaBe. Fig.~2 shows the structure for the model based on $Z_0$.
Since the potential is computable to high precision, one can 
follow the evolution of the vacua for growing $W_0$ into a stringy regime at smaller volumes, 
where the supersymmetric vacuum of \KKLT\ at large $t$ ceases to exist 
and is replaced by a non-supersymmetric vacuum. 
It would be interesting to study this further and we plan to give a detailed discussion elsewhere \BMip.

\subsec{Strongly coupled IR physics and  $\cx N=1$ Minkowski vacua}
The flux potential \wf\ allows for a large class of $\cx N=1$ supersymmetric Minkowski ground states 
of the effective $\cx N=2$ supergravity with $W_0=0$ at the minimum \TriTri\FF.\foot{There are
claims in the literature that supersymmetric solutions in type IIB require the presence of 
negative sources for the space-time tadpole. The type IIB compactifications associated with the 
perturbative vacua 
discussed in this section are well-known counter examples to these claims.}
The condition $W_0=0$ means that there are linear combinations of the period vector $\Pit^\La$,
that vanish at the minimum
\eqn\vp{
v^\al_\La\Pit^\La=0, \qquad \ \Pit^\La=
\pmatrix{-1\cr S U -\h C_a C_a \cr S\cr U \cr C_a}\ .
}
Similarly such null vectors may also exist in vacua with a total $W_0\neq0$ at the minimum.

There are two distinct cases depending on whether one of the null vectors 
is an integral vector of norm $(v^\al)^2=-2$ or not.
In this case there is a classical gauge symmetry enhancement in 8d
from colliding 7-branes.\foot{See \aspe\ for a discussion and further references.}

In the latter case, let us assume that the matter spectrum of the 4d $\cx N=2$ gauge theory from the 7-branes 
leads to an asymptotically free theory.
As is well-known \SW, this theory is strongly coupled near the origin of the Coulomb branch, 
for any non-zero bare coupling. In the present case, the modulus
$\TT$ sets the bare coupling for the 7-brane gauge groups while the positions of the 7-branes
are controlled by the moduli $t_a'$. Tuning the $t_a'$ to obtain an asymptotically free 
4d gauge theory on a subset of 7-branes leads to the divergence of the instanton expansion $\ffi$.

This divergence arises for any large but finite $\Im(\TT)$ from the divergence of the 
coefficient functions $f_k(q_a')$ in \defWi. It follows that one can not use the 
approximation $W=W_F+\cx O(Q)$ in this case; instead one has to resum the instanton series
into a convergent series adapted to this regime, such as in \PPl\PMssb.

Let $\MM$ denote a neighborhood of the subloci in the moduli where there are vectors $v^\al_\La$
of the special type described above associated with an asymptotically free non-Abelian  gauge symmetry.
The theory is strongly coupled in this regime and the subspace $\MM$ should therefore be deleted
from the moduli space when studying the approximation to the classical superpotential $W=W_F+\cx O(Q)$.%
\foot{This holds also in toroidal orientifold compactifications which have sectors
with asymptotically free $\cx N=2$ supersymmetric gauge theories. These orientifolds are strongly coupled
and destabilized by the above effects from gauge instantons.} There may also be vacua in the
strongly coupled region $\MM$, however these are not small deformations of vacua of the classical
potential $W_F$, and need to be studied by the different means of a strong coupling expansion.

These statements hold independently of whether the value $W_0$ of the superpotential 
at the minimum is zero or not. However we will now argue that in the supersymmetric case $W_0=0$,
the weakly coupled vacua are unstable against decompactification and thus artifacts of the 
8d approximation. We will only sketch the argument here, pointing out the general structure and
the two branches of solutions. A more detailed account can be found in \PMssb,
in the context of heterotic string compactifications. The gravitino mass matrix 
in the effective $\cx N=2$ supergravity is 
$$
S_{AB}\sim \pmatrix{-\bb P_\La X^\La & P_\La^3 X^\La\cr P^3_\La X^\La & P_\La X^\La},
$$
where $P^x_\La$, $x=1,2,3$ are the Killing prepotentials describing the flux. In particular
$P_\La=P^1_\La+i P^2_\La$ is the complex combination related to the superpotential \fres\ and
$P^3$ enters the $D$-term \Dterm.
Near a $\cx N=1$ vacuum with $dW=W=0=V_D$, the mass matrix is of the form
\eqn\scorr{
S_{AB}\sim \pmatrix{-W_0+2i(P^2_\La X^\La)_0&0\cr0&W_0}+\pmatrix{-\bb p_0\, \ffi&0\cr0&p_0\, \ffi},
}
where $\ffi$ is the D3-instanton correction defined in \Frho\ and a subscript 0 denotes
the vacuum value. By assumption, $W_0=0$ in the original vacuum and 
the left upper corner is non-zero due to the extra term;
otherwise one would have two massless gravitinos and $\cx N=2$ supersymmetry.

The second term in \scorr\ is the instanton correction, where $p_0$ denotes the component of $P_\La$
multiplying the instanton corrected period.\foot{This is $P_2$ in the notations of eq.\newbas.}
One might expect that for a small correction $\ffi$, the moduli in $W_0$ may adapt themselves to 
restore a zero eigenvalue by a small shift in $W_0$. However this hope is excluded by a
no-go theorem \CGP. This no-go theorem states that whenever the symplectic section $(X^\La,F_\La)$
derives from a prepotential, that is $F_\La=\fc{\p}{\p X^\La}\cx F(X^\La)$, there can not be partial 
supersymmetry breaking to $\cx N=1$ with $dW=W=0$. The section $(X^\La,F_\La)$ in \newbas, can not
be derived from a prepotential if $\ffi=0$ since it does not depend on $\TT$. On the other hand
there is a prepotential after including the $\TT$ dependent 
instanton corrections.

The instanton correction $\ffi$ vanishes at $\TT=i\infty$, where one recovers the original
$\cx N=1$ supersymmetric vacuum. However this vacuum is strictly 8 dimensional and should therefore not 
be counted as a point in the space of 4d string vacua.
The only way to avoid the conclusion that the vacuum decompactifies is 
that there is a new minimum created by the the correction $\ffi$, which means
that the instanton effects are large and  such a vacuum is not described
by the classical superpotential $W_F$.

Thus including the instanton corrections leads to a quite dramatic change in the 
statistics  of perturbative $\cx N=1$ Minkowski vacua, compared to the study of the classical
potential. Most vacua are destabilized by the 
D3 instanton effects and decompactify to eight dimensions. Only
a small fraction is possibly stabilized by strong infrared dynamics,
in the concrete case by confining gauge theory on the brane.\foot{More generally, 
the strongly coupled infrared phase may also describe less conventional
physics, perhaps conformal fixed points or tensionless strings.} 
Thus including the instantons leads to a new picture with less
vacua, which moreover are all ``semi-realistic'' in the (weak) sense
that they lead to interesting low energy physics.

\vskip10pt
\noi {\bf Acknowledgments:} We thank 
R. Minasian,
J. Louis,
S. Kachru,
D. L\"ust
and P. Tripathy 
for valuable discussions and correspondence. 
PB would like to thank the organizers of the String Phenomenology workshop at the Perimeter Institute for a stimulating environment
where some of this work was done.
PB is supported by NSF grant PHY-0355074 
and by funds from the College of Engineering and Physical Sciences  at
the University of New Hampshire.
PM is supported by the German Research Foundation (DFG).

\appendix{A}{Calabi-Yau geometries}
In this appendix we collect some computations in the dual closed string \CY\ manifolds.
The \CY\ manifold $Z$ will be represented as a toric hyper surface, defined as the zero of 
a homogeneous polynomial $p_Z(x_i)=0$ in $n$ variables $x_i$ defined on the space
$(\IC^n\backslash \Xi)/(C^*)^k$. Here $\Xi$ is a certain set removed from $\IC^n$. The
$k$ $\IC^*$ actions act on the variables by multiplication $x_i\to x_i\la^{l_i^{(k)}}$,
with $\la\in\IC^*$ and $l_i^{(k)}$ some integers. The simplest
non-trivial example is $\IP^n$ with $\Xi=\{x_i=0,\ \forall i\}$ and $l_i^{(k)}=1\ \forall i$.
The data of $Z$ can be represented by a convex polyhedron $\Delta$ in the real extension of an
integral lattice isomorphic to $\IZ^4$.
See \TOR\ for background material on toric geometry, 
\mirrlec\ for a review of the computation of
world-sheet instantons in this framework
and \hpkatz\ for many useful informations and links.

\subsec{2 D7-branes}
A minimal choice for an elliptic and K3 fibered manifold $Z$ with three K\"ahler moduli $t_a$ 
is an elliptic fibration over the Hirzebruch surfaces $F_n$ with $n=0,1,2$.\foot{Fibrations
with $n>2$ describe gauge backgrounds that leave more than two independent 7-brane moduli on \ktk, 
as follows from F-theory/heterotic/F-theory duality and the results of \MV.}
Let $Z_n$ denote the elliptic fibration over $F_n$. The manifold $Z_n$ is described by a toric polyhedron 
spanned by the 7 vertices
\eqn\vertiii{
[0,0,-1,0],[0,0,0,-1],[0,0,2,3],[1,n,2,3],[-1,0,2,3],[0,1,2,3],[0,-1,2,3].}
Each vertex is associated with a coordinate $x_i$ and a divisor $D_i:\, x_i=0$ in the
standard way \TOR.

The classical prepotential depends on the topological data as described in eq.\cypp.
All manifolds have $\chi=-480$, whereas the other data vary:
$$\eqalign{
Z_0:\ c_2(J_a)=(92,24,24),\ \ \cx F_{cubic}&=\fc{4}{3} t_1^3+t_1^2 t_2+t_3\, (t_1 t_2+t_1^2),\cr
Z_1:\ c_2(J_a)=(92, 36, 24),\ \ \cx F_{cubic}&=\fc{4}{3} t_1^3+\fc{3}{2} t_2 t_1^2+\fc{1}{2} t_2^2 t_1
+t_3\,  (t_2 t_1+t_1^2),\cr
Z_2:\ c_2(J_a)=(92, 48, 24),\ \ \cx F_{cubic}&=\fc{4}{3}t_1^3+2 t_2 t_1^2+t_2^2 t_1+t_3\,(t_1t_2+ t_1^2).
}
$$
The manifold $Z_1$ has another, flopped geometric phase, see e.g. \KMV. By a change of variables 
$t_1=U,t_2=T-U,t_3=S-(1-\fc{n}{2})\, U-\fc{n}{2}\,T$, 
the cubic parts of the  prepotential can be be brought into the standard form $\cx F=STU+\fc{1}{3}U^3$,
which is the prepotential in the perturbative heterotic string variables \LSTY.

The three K\"ahler moduli $t_a$ are the volumes of the curves in the classes $C_a$:
$$
C_1=D_4.D_6,\ \ C_2=D_3.D_4,\ \ C_3=D_3.D_6,
$$
that represent the elliptic fiber $E$, the fiber $F$ of $F_n$ and the exceptional section $B$
of $F_n$, respectively. The variable $t_3$ enters linearly in the prepotential. We identify it with the
volume of the lower K3 $\kb$, $t_3=\TT$ and use $Q$ to denote the D3-instanton weight \instw.
The two other moduli describe the positions of two independent
D7-brane on base $\IP^1$ of the upper K3 $\kf$ in \ktk.

\subsec{2 D7 branes and 1 D3 brane}
The manifold is described by a toric polyhedron spanned by the 8 vertices
$$\scriptstyle
[0,0,-1,0],[0,0,0,-1],[0,0,2,3],[1,1,2,3],[-1,0,2,3],[0,1,2,3],[0,-1,2,3],[-1,1,2,3]\ .
$$
There are four geometric \CY\ phases related by flops. We consider a phase where the 
D3-brane is in a generic position.
The toric diagram of the base and the triangulation are shown in Fig.~3. 
\vskip 0.5cm
{\baselineskip=12pt \sl
\goodbreak\midinsert
\centerline{\epsfxsize 1  truein\epsfbox{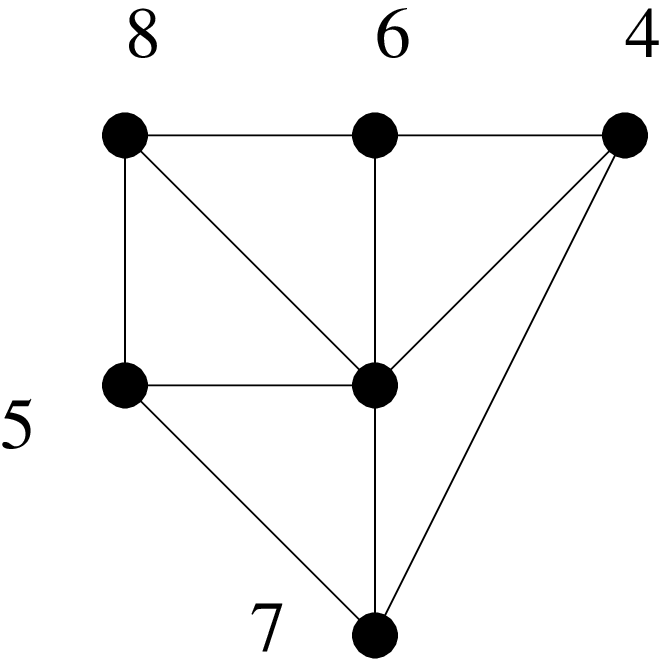}}
\leftskip 1pc\rightskip 1pc \vskip0.3cm
\noindent{\ninepoint  \baselineskip=8pt 
{{\bf Fig.~3:} Triangulation of the $F_1$ base.}}\endinsert}
\noi 
The cubic part of the  prepotential is
\eqn\Fmodii{
\cx F_{cubic}= (t_2 t_1+t_3 t_1+t_1^2)\ t_4+(2 t_3 t_2 t_1+\fc{3}{2} t_2 t_1^2+\fc{7}{6} 
t_1^3+\fc{1}{2} t_2^2 t_1+t_3^2 t_1+2 t_3 t_1^2)\ ,
}
and topological data $\chi=-420$ and $c_2(J_a)=(82, 36, 48, 24)$. 
The four K\"ahler moduli $t_a$ are the volumes of the curves in the classes $C_a$:
$$
C_1=D_4.D_6,\ \ C_2=D_3.D_5,\ \ C_3=D_3.D_8,\ \ C_4=D_3.D_6.
$$
$C_1$ is the class of the elliptic fiber, $C_2$ is the proper transform of the fiber $F$
of $F_1$ after the blow up, $C_3$ is the exceptional $\IP^1$ 
and $C_4$ is the class of the section of the
K3 fibration.

The translation to moduli in the dual F-theory is as follows. 
The volume of $C_4$, $t_4$, is again identified with the volume modulus 
$\TT$ of the dual K3 $\kb$. The three other moduli correspond to two 
independent D7 brane moduli, and one D3-brane modulus. The two moduli $t_2$ and 
$t_3$ describe the position of one D3-brane relative to the 7-branes and
encode therefore already one position of the two 7-branes. One can 
take $t_2$ or $t_3$ as the position of the D3-branes, while $t_1$ and $t_3+t_2$
as the two positions of the two D7-branes; in fact $t_3+t_2$ is the 
volume of the curve $D_3.D_4$ which is the section of the (elliptic fibration of the) K3 fiber
which has already been identified with one of the D7-moduli in the manifold $Z_1$ without
D3-brane.

There is a single light hyper multiplet from 37 strings for $t_2=0$ or $t_3=0$.
There is another hyper multiplet with mass $t_3+t_4$ which gets 
light at strong coupling or small K3 volume. 
As explained before, each hyper multiplet is the first state in an infinite tower of 
particles associated with a non-critical string with $E_8$ current algebra.
This is confirmed by a computation of the Gromov-Witten invariants
\eqn\ncs{
\ffi=q_*\, (1+252q_1+5130q_1^2+54760q_1^3+...)=q_*\, \fc{E_4}{\eta^{12}}(q_1),
}
for $q_*\in \{q_2,q_3,q_3q_4\}$.
In fact the divisors $D_5$ and $D_8$ are described locally by a Weierstrass equation 
$$
y^2+x^3+xzf_4+z^6g_6=0\ ,
$$
in $\IC^5/(\IC^*)^2$, where subscripts denote the homogeneous degree of the polynomials
$f$ and $g$ on the coordinates on a $\IP^1$ base. The sum in \ncs\ arises from 2-cycles in 
that wrap once the section and $k$ times the elliptic fiber. 

The instanton correction to the superpotential is contained in the period $\p_4\cx F$. 
{}From the prepotential \Fmodii, and the special choice of section eq.\newbas, 
it follows that the classical superpotential
does not depend on the moduli $t_4$ and $t_2-t_3$. As expected the two unfixed
moduli in the dual theory are the K3 volume and the D3-brane position, whereas
the D7-brane positions $t_1$ and  $t_2+t_3$ are fixed by the classical superpotential.

Here are two possible parametrizations of the moduli adapted to the F-theory dual.
The first is 
\eqn\pari{\eqalign{
&t_1=\SS,\ t_2=\fc{1}{2}(U-\SS)-\BB,\ t_3=\fc{1}{2}(U-\SS)+\BB,\ t_4=\TT-\fc{7}{8}U-\fc{1}{2}\BB,\cr
&\cx F=\ \TT \SS U -\fc{1}{2}\SS \BB^2 +\fc{7}{24}\SS^3 + (-\TT-\fc{7}{8}U-\fc{5}{3}\SS).
}}
This corresponds to a D3-brane position with $\BB=0$ centered away from the 7-branes, valid in the patch
$\Im t_a>0$. The shifts in $\TT$ have been chosen to simplify the cubic terms.
The second parametrization is 
\eqn\pari{\eqalign{
&t_1=\SS,\ t_2=\BB,\ t_3=U-\SS-\BB,\ t_4=\TT-U,\cr
&\cx F=\ \TT \SS U -\fc{1}{2}\SS \BB^2  -\fc{1}{2}\SS^2 \BB +\fc{1}{6}\SS^3 + (-\TT-U-\fc{17}{12}\SS+\fc{1}{2} \BB).
}}
This corresponds to a D3-brane sitting on top of one of the 7-branes for $\BB=0$.

Note that in both cases, $\Im t_a>0$ requires $U>\SS$.

\subsec{3 D7-branes and 1 D3 brane}
For the identification of the effective supergravity theories it is 
interesting to consider one extra D7 brane.
We consider the manifold $Z_0^\flat$ defined by the vertices
$$\scriptstyle
[[0, 0, -1, 0], [0, 0, 0, -1], [0, 0, 2, 3], [1, 0, 2, 3], [-1, 0, 2, 3], 
[0, 1, 2, 3], [0, -1, 2, 3], [1, 1, 2, 3], [0, -1, 1, 2]].
$$
This manifold is related to $Z_0$ by a blow up in the K3 fiber, corresponding to a further independent
D7-brane, and a blow up in the base $F_0$ corresponding to one space-filling D3-brane.\foot{This
\CY\ manifold has also been studied in \LSTY\ as model \#16.} There are several geometric CY phases in the
K\"ahler moduli space. We consider one, where the D3-brane is at a generic position. 
The topological data are $\chi=-360$, $c_2(J_a)=(136,82,24,24,36)$ and 
$$\eqalign{
\cx F_{cubic}=
&(3 t_1^2+t_2^2+2 t_1 t_5+4 t_2 t_1+t_2 t_4+t_2 t_5+2 t_1 t_4) t_3+\cr
&12 t_2 t_1^2+\fc{1}{2} t_5^2 t_2+6 t_1 t_5 t_2+7 t_1 t_2^2+t_2 t_4 t_5+
4 t_4 t_1^2+\fc{3}{2} t_5 t_2^2+t_4 t_2^2+\cr
&\fc{7}{6} t_2^3+\fc{20}{3} t_1^3+t_1 t_5^2+
5 t_5 t_1^2+4 t_2 t_4 t_1+2 t_1 t_5 t_4\ .
}$$
There are two linear variables, $t_3,t_4$  that can be identified with the K3 volume $\TT$. 
This corresponds to two dual compactifications of F-theory on \ktk.  
With the parametrization
$$\eqalign{
&t_1=C,\  t_2=\SS-2C,\  t_3=\TT-\fc{1}{2}U-\fc{3}{2}\SS-\BB,\cr 
&t_4= \fc{1}{2}U-\fc{1}{2}\SS-\BB,\  t_5=\fc{1}{2}U-\fc{1}{2}\SS+\BB,
}$$
the prepotential becomes
$$\Fc=
\TT\, (-C^2+\SS U)+(-\fc{1}{2} \BB^2 \SS-\SS^2 U-\fc{1}{2} \SS U \BB+\fc{7}{24} \SS^3+\fc{4}{3} C^3-\fc{1}{8} \SS U^2)\ .
$$
\listrefs
\end